\documentclass[lettersize,journal]{IEEEtran}

\def\Bc{{\mathcal B}}

\def\Gc{{\mathcal G}}

\def\Gbf{{\mathbf G}}

\def\Jc{{\mathcal J}}

\def\Lc{{\mathcal L}}

\def\Mc{{\mathcal M}}

\def\Rbb{{\mathbb R}}

\def\0{{\bf 0}}

\def\underdotline#1{{\setbox0=\hbox{#1}\rlap{\raise-0.28em\hbox to\wd0{\rm\tiny\cleaders\hbox{.\kern-0.1ex}\hfill}}\box0}}
\def\underdashline#1{{\setbox0=\hbox{#1}\rlap{\raise-0.28em\hbox to\wd0{\rm\tiny\cleaders\hbox{-\kern-0.1ex}\hfill}}\box0}}
\def\understarline#1{{\setbox0=\hbox{#1}\rlap{\raise-0.28em\hbox to\wd0{\rm\tiny\cleaders\hbox{$\star$\kern-0.1ex}\hfill}}\box0}}

\newcommand{\bitem}{\begin{itemize}}
\newcommand{\eitem}{\end{itemize}}
\newcommand{\btabular}{\begin{tabular}}
\newcommand{\etabular}{\end{tabular}}
\newcommand{\bcenter}{\begin{center}}
\newcommand{\ecenter}{\end{center}}
\newcommand{\bea}{\begin{eqnarray}}
\newcommand{\eea}{\end{eqnarray}}
\newcommand{\bean}{\begin{eqnarray*}}
\newcommand{\eean}{\end{eqnarray*}}
\newcommand{\ba}{\left[ \begin{array}}
\newcommand{\ea}{\\ \end{array} \right]}
\newcommand{\bear}{\begin{array}}
\newcommand{\eear}{\\ \end{array}}

\newcommand{\non}{\nonumber}

\font\myownfont=cmr17 scaled \magstep5
\def\psfancypar#1#2{\def\biginitial#1{{\myownfont#1}}%
  \def\makeinitial#1{\setbox8\hbox{\strut\vbox to 1.3ex
    {\hbox{\biginitial#1}\vskip -4pc plus 3.5pc minus 3.5pc}}}%
  \makeinitial#1%
  \ifdim\parindent>1.3\wd8\dimen8=\parindent
     \else\dimen8=1.3\wd8\fi
  \hangindent=\dimen8\hangafter=-2
  \noindent
  \strut\hskip-1\dimen8\box8{\sc#2}}%

\newcounter{subequation}
\def\beasub{\addtocounter{equation}{+1}
\setcounter{subequation}{\value{equation}}
\setcounter{equation}{0}
\renewcommand{\theequation}{\arabic{subequation}\alph{equation}}
\begin{eqnarray}}
\def\eeasub{\end{eqnarray}
\setcounter{equation}{\value{subequation}}
\renewcommand{\theequation}{\arabic{equation}}}

\newcommand{\overbar}[1]{\mkern 1.5mu\overline{\mkern-1.5mu#1\mkern-1.5mu}\mkern 1.5mu}

\newcommand{\bos}{\boldsymbol}

\newcommand{\bbm}{\begin{bmatrix}}
\newcommand{\ebm}{\end{bmatrix}}
\newcommand{\abs}[1]{\lvert #1 \rvert}

\usepackage{amsmath,amsfonts} 
\usepackage{xcolor}
\usepackage{amssymb}
\usepackage{url}
\usepackage{graphicx}
\usepackage[export]{adjustbox}
\usepackage{array}
\usepackage{threeparttable}
\usepackage{multirow}
\usepackage{algorithmic}
\usepackage{mathtools}
\usepackage{algorithm}
\usepackage[shortlabels]{enumitem}
\usepackage[utf8]{inputenc}
\usepackage{textcomp}
\usepackage{stfloats}
\usepackage{url}
\usepackage{verbatim}
\usepackage{graphicx}
\usepackage{cite}
\usepackage[caption=false,font=footnotesize,labelfont=sf,textfont=sf]{subfig}
\usepackage{amsthm}
\usepackage{array,multirow}
\usepackage{multicol}

\newtheorem{theorem}{Theorem}

\newtheorem{rmk}{Remark}
\newtheorem{assumption}{Assumption}
\newtheorem{definition}{Definition}

\newcommand{\pii}[1]{p_{#1,i}} 
\newcommand{\bpi}[1]{{\tilde{p}}_{#1,i}}
\newcommand{\btp}[1]{{\bar{p}}_{#1,i}}
\newcommand{\xii}[1]{{#1}_{i,i}}
\newcommand{\ix}[2]{{#1}_{#2,i}}

\begin{document}
	
	\title{Distributed MPC for autonomous ships on inland waterways with collaborative collision avoidance}

    \author{Hoang Anh Tran, Tor Arne Johansen, Rudy R. Negenborn
        \thanks{This work has been submitted to the IEEE for possible publication. Copyright may be transferred without notice, after which this version may no longer be accessible.}
		\thanks{The research leading to these results has received funding from the European Union's Horizon 2020 research and innovation programme under the Marie Skłodowska-Curie grant agreement No 955.768 (MSCA-ETN AUTOBarge), and the Researchlab Autonomous Shipping at Delft University of Technology. This publication reflects only the authors' view, exempting the European Union from any liability. Project website: http://etn-autobarge.eu/.}
		\thanks{Hoang Anh Tran and Tor Arne Johansen are with the Department of Engineering Cybernetics, Norwegian University of Science and Technology (NTNU), Norway (e-mail: hoang.a.tran@ntnu.no, tor.arne.johansen@ntnu.no).}
		\thanks{Rudy R. Negenborn is with the Department of Maritime and Transport Technology, Delft University of Technology, The Netherlands (e-mail: r.r.negenborn@tudelft.nl)}} 
	\markboth{Journal of \LaTeX\ Class Files,~Vol.~14, No.~8, August~2021}%
	{Shell \MakeLowercase{\textit{et al.}}: A Sample Article Using IEEEtran.cls for IEEE Journals}
	
	\maketitle
\begin{abstract}
This paper presents a distributed solution for the problem of collaborative collision avoidance for autonomous inland waterway ships.
A two-layer collision avoidance framework that considers inland waterway traffic regulations is proposed to increase navigational safety for autonomous ships.
%
%
%
Our approach allows for modifying traffic rules without changing the collision avoidance algorithm, which is based on a novel formulation of model predictive control (MPC) for collision avoidance of ships.
This MPC formulation is designed for inland waterway traffic and can handle complex scenarios.
The alternating direction method of multipliers (ADMM) is used as a scheme for exchanging and negotiating intentions among ships.
Simulation results show that the proposed algorithm can comply with traffic rules.
Furthermore, the proposed algorithm can safely deviate from traffic rules when necessary to increase traffic efficiency in complex scenarios.
\end{abstract}




\maketitle









\section{Introduction}\label{intro}
\subsection{Background and motivation}
Developing inland autonomous surface ships is receiving increasing attention over the past years.
The key to allowing the operation of an autonomous ship in inland waterway traffic (IWT) is the guarantee of navigation safety.
A collision avoidance system (CAS) should ensure navigation safety of an autonomous ship.
Different approaches for the CAS of inland autonomous ships have been proposed, including MPC \cite{mahipala_model_2023}; Velocity obstacle \cite{zhang_22}; Potential field method \cite{gan_22,yan_20}; and Scenario-based MPC (SB-MPC) \cite{tran_collision_2023}.
Additional information about different solutions for developing CASs for inland autonomous ships can be found in \cite{tran_survey_2023}.

When it comes to CASs, one of the challenges is predicting the intention and future trajectories of neighboring ships.
Conventionally, a CAS predicts future trajectories of neighboring ships based on the information from onboard sensors and the AIS system.
The current development of digital communication technology allows route exchange and intention sharing among ships \cite{monalisa,guiking_digital_nodate}.
An intention exchange system combined with a distributed computing framework opens opportunities for collaborative CAS (C-CAS) of autonomous ships.

Over the past decade, proposals have been made to synthesize collaborative collision avoidance controllers for autonomous ships \cite{akdag_collaborative_2022,zheng17,du_22}. 
%
%
Frameworks for the C-CAS problem can be categorized as centralized or distributed.
On one hand, a centralized framework uses one central computing unit to calculate a solution for C-CAS for all ships and broadcasts the solution over a network \cite{chen_18,tam13,kurowski_19}.
With this approach, the collision avoidance solution is usually globally optimal and does not require computational resources on each ship.
However, this approach lacks scalability and robustness \cite{akdag22}.
On the other hand, in a distributed framework, all ships within an area will calculate their own collision avoidance solutions and negotiate among one another through communication to adjust their own solutions \cite{chen18,kim_2017,tang_22}.
Although the distributed C-CAS usually offers a locally optimal solution, robustness and scalability properties can lead to a distributed approach being a more promising solution for the C-CAS problem.

%
Because of the recent improvements in solvers for optimization problems, MPC has seen increasing applications in autonomous vehicles \cite{eriksen_20,schwarting2017parallel,kneissl_18}.
MPC has shown the ability to significantly increase autonomous vehicles' safety even in complex traffic scenarios \cite{yu_model_2021}.
In order to apply MPC to a distributed C-CAS problem, the Alternating Direction Method of Multipliers (ADMM) is one of the well-known solutions used.
Several studies have exploited the ADMM scheme to address the C-CAS problem for inland autonomous ships.
A distributed MPC scheme approach based on ADMM was proposed by \cite{zheng17} to deal with the C-CAS problem in intersection crossing situations.
The proposed communication scheme there, however, is not fully distributed since it requires one ship to take the coordinator role.
A fully distributed nonlinear MPC-based ADMM, which requires no coordinator, is presented in \cite{ferranti18}.
Another fully distributed MPC scheme is proposed by \cite{chen18} to deal with the C-CAS problem for vessel train formations.
Unlike other ADMM schemes, \cite{chen18} uses a serial iterative ADMM scheme instead of a parallel ADMM scheme.
The iterative scheme allows ships to exploit up-to-date information from neighboring ships and is more suitable for transport networks \cite{negenborn14}.
In \cite{chen18}, how the update order for each ship within the serial iterative ADMM scheme affects the behavior of ships is also discussed, although a specific order is not suggested.

Although significant effort has been devoted to the problem of CAS for inland autonomous ships, none of the existing approaches explicitly consider the IWT regulations.
In the existing literature, algorithms either consider a limited subset of the \textit{Convention on the International Regulations for Preventing Collision at Sea} (COLREGS) or do not comply with any traffic regulations.
Besides, methods considering COLREGS cannot be used directly, as different rules regulate inland waterways.
To be allowed to operate in inland waterway traffic, ship's behaviors should comply with the local traffic regulations.
Furthermore, it is recommended that, in case one ship needs to take action to avoid a collision, the altering of course or speed should be large enough to be readily apparent by neighboring ships \cite{binnen,ccnr}.
One of the main obstacles when applying inland waterway regulations to collision avoidance algorithms is that the IWT regulations can change depending on the region of the waterways.
For instance, the Netherlands' waterway is regulated by \textit{Binnenvaartpolitiereglement} \cite{binnen}, while \textit{Police regulations for the navigation of the Rhine} regulate the Rhine River \cite{ccnr}.
A ship voyage could start in one region and end in another, meaning that the IWT regulation may change during the voyage. 
This poses a challenge for designing the CAS since changing the traffic rules could result in changing the behavior of the CAS algorithm.

\subsection{Proposed two-layer framework for collision avoidance} \label{2layer}
\begin{figure}[!b]
	\centering
	\includegraphics[width=1\linewidth]{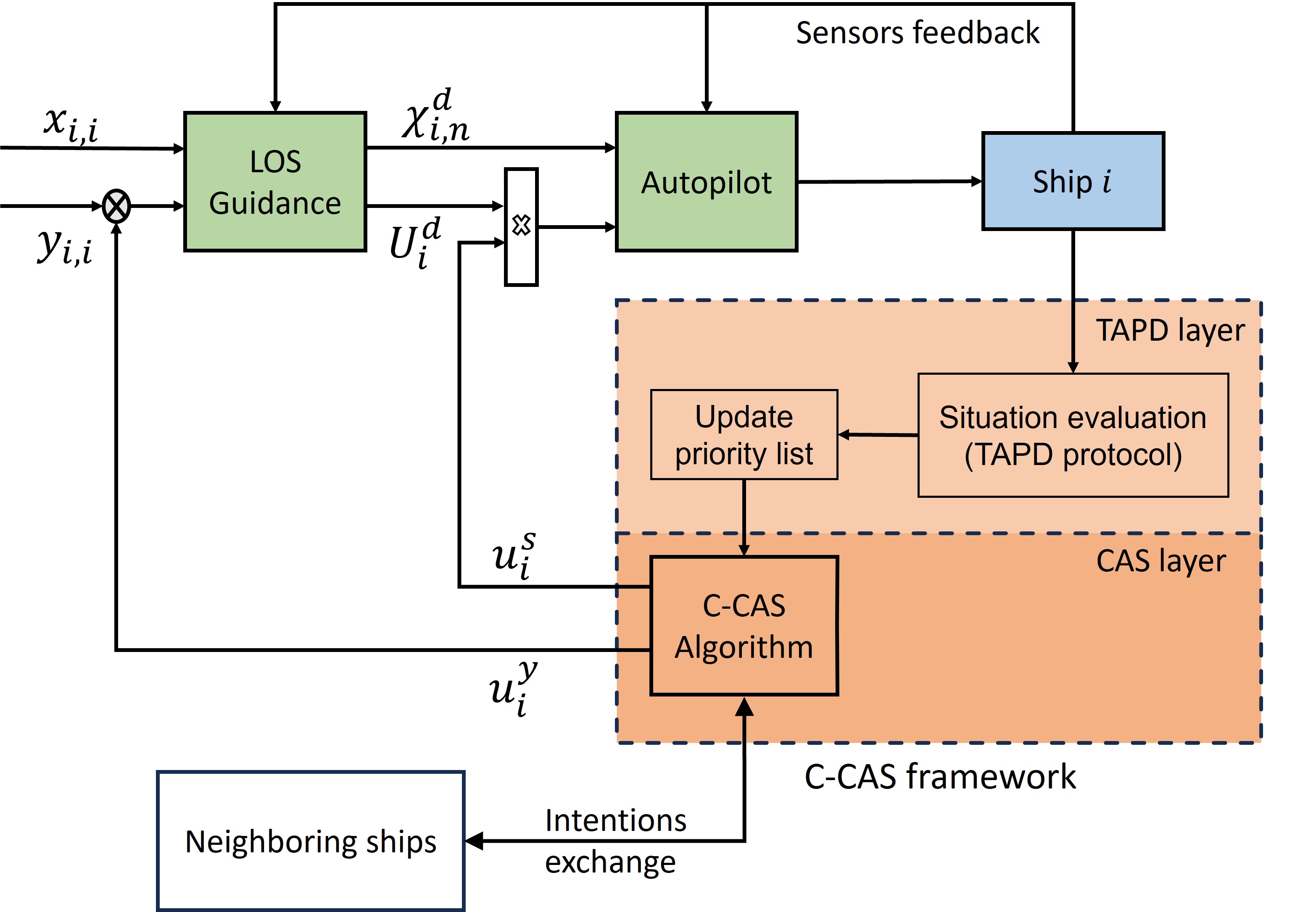}
	\caption{Control scheme with the proposed C-CAS framework: \textit{ $U_i^d$ and $\chi_{i,n}^d$ are nominal surge speed and course angle command (in the inertial frame $\{n\}$), respectively, used as input to an autopilot that contains course and speed control loops. The control signals from CAS are cross-track offset, $u^y_i$, and speed modification, $u^s_i$}.}
	\label{CAS-protocol}
\end{figure}

In this paper, we introduce a C-CAS framework to deal with the inland waterway traffic rules compliance.
The C-CAS framework includes two layers (see Fig. \ref{CAS-protocol}): the first layer is the traffic assessment \& priority determination protocol (TAPD), and the second layer is the C-CAS algorithm.
In the first layer, the traffic situation will be evaluated, and based on traffic rules a priority list is created through the communication protocol.
The priority list is then used to determine the order of executing the C-CAS algorithm for each ship.
%
%
%
%
%

Following the inland traffic rules, in encountering situations between two ships, e.g., head-on, overtaking, and crossing, one ship is allowed to stand-on, and the other must give-way.
A cost function can introduce the stand-on and give-way behavior using a binary variable as in \cite{johansen16,tengesdal_22}.
However, this method is only effective for discrete input optimization algorithms such as SB-MPC. 
One way to apply this approach in MPC-based algorithms is to approximate the binary variable with a smooth function \cite{eriksen_20}.
Nonetheless, this approximation could increase the complexity of the MPC problem.
Moreover, the traffic rules could be changed depending on the region in which a ship is sailing.
If the traffic rules are integrated inside the CAS algorithm, then the CAS algorithm also has to be reformulated whenever the traffic rules change.
%


The advantages of our C-CAS framework are twofold. 
Firstly, the framework limits the complexity of the C-CAS algorithm by shifting the traffic rules out of the optimal control problems.
Secondly, we can change the traffic rules by adjusting the upper layer, i.e., TAPD protocol, without changing the C-CAS algorithm.

\subsection{Contributions}
This paper proposes a novel C-CAS framework and  a distributed  algorithm that explicitly considers IWT regulations.
The algorithm is based on distributed MPC (DMPC) using the ADMM framework.
Each vessel individually computes a collision avoidance solution and exchanges intentions with neighboring ships through a hierarchical protocol.
The main contributions of this paper are as follows:
\begin{enumerate}
	\item A new approach to the problem of traffic rules compliance for collision avoidance: By introducing a two-layer framework, we separate the task of ensuring traffic rule compliancy and avoiding collisions at different layers. This separation allows for modifying traffic rules (e.g. based on the region) without reformulating the CAS algorithm.
	\item A DMPC approach addresses the CAS problem for inland autonomous ships: This DMPC approach is an improvement of the algorithm presented in \cite{tran_collision_2023}. A risk function is proposed to evaluate risks of collision between ships. By minimizing the risk function in the DMPC problem, we increase the navigation safety of ships. Besides, we use the serial iterative ADMM scheme to achieve a solution for the collaborative collision avoidance problem. Unlike the algorithm presented in \cite{chen18}, which is only suitable for bi-directional waterway scenarios, our algorithm can also solve intersection crossing scenarios.
	\item The behaviors of ships using the proposed algorithm comply with traffic rules for the case of two-ship encounters. Most existing regulations only take this situation explicitly into account. In addition to this, the proposed algorithm reduces the collision risk in case of more than two ships.
	\item The proposed algorithm is verified in simulation experiments with various representative traffic scenarios.
\end{enumerate}
\subsection{Outline}
The remainder of this paper is organized as follows. 
Section \ref{pre} provides preliminaries on the coordinate systems, notation, and nonlinear ADMM.
%
Section \ref{sec:comm} presents the communication protocol adopted in the first layer.
Section \ref{sec:colab} describes the DMPC approach for the C-CAS problem adopted in the second layer.
The results of simulation experiments are presented in Section \ref{sec:sim}, and Section \ref{sec:cons} concludes and provides directions for future research.
\section{Preliminaries} \label{pre}
\begin{figure}[!t]
	\centering
	\includegraphics[width=0.8\linewidth]{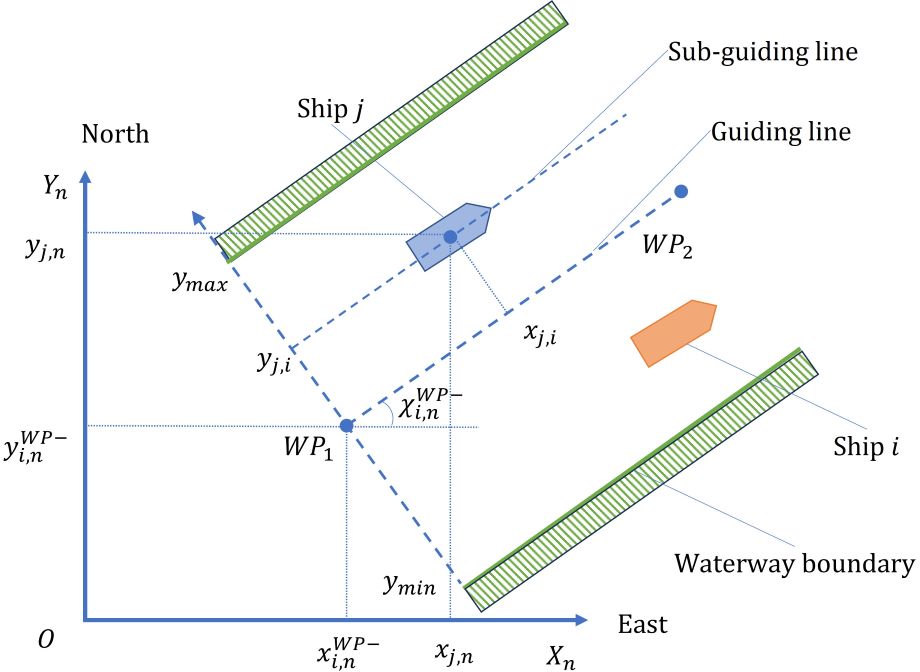}
	\caption{Path coordinate and inertial coordinate.}
	\label{coordinate}
\end{figure}
\subsection{Notation}
We denote the set of $M$ ships as $\Mc$, and use the term "ship \textit{i}`` to refer to the ship with index $i$.
The state of ship $j$ in the inertial frame $\{n\}$ is $\eta_j = [x_{j,n},y_{j,n}, \chi_{j,n}]^\top$, where $x_{j,n},y_{j,n}$ are the position of the ship in $X_n$, $Y_n$ axis and $\chi_{j,n}$ is the course angle, i.e., the angle from the $X_n$ axis to the velocity vector of the ship  (see Fig. \ref{coordinate}).
The velocity vector of ship $j$ in the inertial frame $\{n\}$ is denoted as $v_{j,n}$.
We use the path coordinate frame $\{w_i\}$ to illustrate the position of ship $j$ along the defined waypoints of ship $i$ as in \cite{zheng16}.
The position of ship $j$ with respect to the path coordinate frame $\{w_i\}$ is defined as:
\begin{align*}
	\pii{j} &=\bbm \ix{x}{j} \\ \ix{y}{j} \\ \ix{\chi}{j} \ebm = R_i \bbm x_{j,n} - x_{j,n}^{WP^-} \\ y_{j,n}- y_{i,n}^{WP^-}  \\ \chi_{j,n} - \chi_{i,n}^{WP^-}  \ebm = R_i (\eta_j-{\eta}_{i}^{WP^-}),\\
	R_i &=\bbm \cos\left(\chi_{i,n}^{WP^-}\right) & \sin\left(\chi_{i,n}^{WP^-}\right) & 0 \\
	-\sin\left(\chi_{i,n}^{WP^-}\right) & \cos\left(\chi_{i,n}^{WP^-}\right)  & 0 \\
	0& 0 & 1\ebm,
\end{align*} 
where ${\eta}_{i}^{WP^-} =[x_{i,n}^{WP^-},y_{i,n}^{WP^-}, \chi_{i,n}^{WP^-}]^\top $ contains the parameters of the previous active waypoint in the inertial frame $\{n\}$ (see Fig. \ref{coordinate}).
We use the notation $\pii{i} = [x_{i,i},y_{i,i}, \chi_{i,i}]^\top$ to denote the state of ship $i$ in the path coordinate frame $\{w_i\}$ of ship $i$.
We denote the line segment that connects waypoints as the guiding line, and refer to a sub-guiding line as a line that is parallel with the guiding line.
Moreover, the extended-real line is $\bar{\Rbb} = \Rbb \cup \{\infty\}$, and the Kronecker product is $\otimes$.
%
%
%
%

\subsection{Assumptions}
We assume that the information from every ship is exchanged perfectly over the network:
\begin{assumption}
	We consider the network of ships as a graph $\Gc = (v, e)$. Then, $v$ is the set of ships; $e$ is the set that represents the communication links between ships. We assume that at each time step $k$, the graph $\Gc$ is an undirected connected graph.
\end{assumption}
\begin{assumption}
	We assume ships exchange information through a synchronous communication protocol, i.e., there is sufficient bandwidth for communication, and there is no delay or packet loss. 
\end{assumption}
\begin{assumption}
	We assume all ships in $\Mc$ exchange their position information continuously. Through communication and sensor information from their own, all ships have the same perception of the traffic situation.
\end{assumption}
\subsection{Nonlinear Alternating Direction Method of Multipliers}
%
In Section \ref{sec:colab} we introduce a distributed MPC algorithm to address the problem of collaborative collision avoidance.
In this approach, we solve a distributed optimization problem to find a solution for ships to avoid collisions.
This distributed optimization is solved cooperatively by each ship's controller that is solving its local optimization problem combined with information exchange between ships.
We use the ADMM algorithm as a scheme for solving cooperative optimization problems and information exchange.
More specifically, we use the Nonlinear Alternative Method of Multipliers (NADMM) that is proposed in \cite{themelis_20} as the basis for our algorithm.
Compared with the original ADMM, this approach does not require the local optimization problem to be convex to ensure the algorithm's local convergence.
Instead, the NADMM exploits the so-called \textit{Douglas--Rachford splitting} to guarantee the tight convergence for nonconvex optimization problems.

Consider the following optimization problem:
\begin{subequations}\label{ref-prob}
	\begin{align}
		\min_{(w,v) \in \Rbb^m \times \Rbb^n} \quad & f(w) + g(v)\label{ref-prob:prob}\\
		\textrm{subject to} \quad  &  \quad  Aw + Bv =b \label{ref-prob:cons},
	\end{align}
\end{subequations}
where $f: \Rbb^m \rightarrow \overbar{\Rbb}$, $g: \Rbb^n \rightarrow \overbar{\Rbb}$, $A \in \Rbb^{q \times m}$, $B \in \Rbb^{q \times n}$, and $b \in \Rbb^q$.
According to \cite{themelis_20}, NADMM iteratively solves problem \eqref{ref-prob} with the following steps:
\begin{align}\label{NADMM}
	\begin{cases}
		z^{+1/2} &= z - \beta(1-\lambda)(Aw + Bv -b),\\
		w^+      &\in \arg\min \Lc_{\beta}(\cdot,v,z^{+1/2}),  \\
		z^{+}    &= z^{+1/2} + \beta(Aw + Bv -b),\\
		v^+      &\in \arg\min \Lc_{\beta}(w,\cdot,z^{+}),
	\end{cases}
\end{align}
in which $\beta>0$, and $\lambda \in (0, 2)$ are penalty and relaxation parameters, respectively. 
Moreover, $\Lc_{\beta}(w,v,z) := f(w) + g(v) + \langle z, Aw + Bv -b\rangle +\frac{\beta}{2}||Aw + Bv -b||^2$ is the augmented Lagrangian, where $z \in \Rbb^q$ is the Lagrange multiplier.
Besides, $(w^+,v^+,z^+)$ denotes the updated state of $(w,v,z)$.
%
%
%

%
%
%
\section{Traffic assessment \& priority determination protocol} \label{sec:comm}
\begin{algorithm}[!b]
	\caption{Priority assignment algorithm on ship $i$} \label{alg:protocol}
	\begin{algorithmic}[1]
		\FORALL{$j \in \Mc$} \label{alg2:sb-assign1}
		\IF{$j$ is on the STARBOARD side of the waterway}
		\STATE{$\rho_{j,i}:=1$}
		\ELSE 
		\STATE{$\rho_{j,i}:=0$}
		\ENDIF \label{alg2:sb-assign2}
		\ENDFOR
		\FORALL{$j \in \Mc, j \ne i$}
		\IF{$T_{i}^{IC} < T_{IC}$\AND $T_{j}^{IC} < T_{IC}$} \label{alg2:crosscon}
		\STATE{\textit{situation$_i$(j)} = CROSSING} \label{alg2:cross-situation}
		\IF{$\rho_{j,i} <1$} \label{alg2:sb-port1}
		\IF{\textit{i} comes from STARBOARD}
		\STATE{$\rho_{j,i} :=\rho_{j,i}+\varpi$}
		\ELSIF {\textit{i} comes from PORT}
		\STATE{$\rho_{j,i} :=\rho_{j,i}-\varpi$}
		\ENDIF\label{alg2:sb-port2}
		\ENDIF
		\ELSIF{\textit{j} HEADON with \textit{i}}
		\STATE{\textit{situation$_i$(j)} = HEADON} \label{alg2:ho-situation}
		\ENDIF
		
		\IF{$\rho_{ii} >\rho_{ij}$}
		\STATE{$p\_list_i(j):=1$}
		\ELSE
		\STATE{$p\_list_i(j):=0$}
		\ENDIF
		\ENDFOR
	\end{algorithmic}
\end{algorithm}
In the first layer of the framework, traffic situation assessment is carried out.
%
%
In the following, the TAPD protocol is presented in Section \ref{sec:comm-prot}.
Then, Section \ref{sec:comm-deadlock} discusses situations where the TAPD protocol could fail and how to resolve these situations. 
\subsection{Protocol development}\label{sec:comm-prot}
From \cite{chen18}, we can observe that when two ships encounter one another, the ship that makes the first decision demonstrates the behavior of the give-way ship and vice versa.
The explanation for this behavior is as follows.
The ship that makes the first decision usually assumes that the other ship keeps the same course and speed.
Assuming that the CAS can make a necessary action to avoid collision, this ship will act as a give-way ship.
The other ship, with the information of CAS's actions from the former ship, can keep its course and portrays the behavior of a stand-on ship.

Influenced by this observation, we propose a protocol to establish the order of making decisions based on priority.
Accordingly, the ship with give-way obligation will get priority to make a decision first, before the ship with stand-on gets permission.
The TAPD protocol is as follows:
\begin{enumerate}
	\item Depending on the situation with other ships and to comply with the traffic rules, every ship $i$ $\left(i \in \Mc\right)$ assigns a relative priority value for all surrounding ships $j$, including itself. 
	We denote the relative priority value between ships $i$ and $j$, assigned by ship $i$ as $\rho_{j,i}$.
	It is not necessarily required that $\rho_{i,i}$ = $\rho_{i,j}$, because the priority values reflect the relation between two ships.
	%
	\item All ships compare their priority values pair-wise to identify their priority in the network. A ship with a lower priority value has to give-way to a ship with higher one.
	\item A ship $i$ is considered to have the highest priority if no other ship has a higher priority value than ship $i$. There may be more than one highest-priority ship.
	\item A ship $i$ is considered to have the lowest priority if no other ship has a lower priority value than ship $i$. There may be more than one lowest-priority ship.
	\item The C-CAS algorithm starts with the lowest-priority ships and ends with the highest-priority ships. In other words, a ship can make a CAS decision when all lower priority ships have made their decisions.
\end{enumerate}

\begin{rmk}
	This protocol is established in a distributed manner, which requires no coordinator.
	Each ship identifies nearby neighboring ships with lower priority and waits until those lower priority ships have updated their decisions.
	Additionally, a ship broadcasts a signal within the network to notify neighboring ships when it has finished its C-CAS algorithm.
	Therefore, a coordinator to store the global priority list is not needed.
\end{rmk}

Our approach considers the rules adopted in the Netherlands' inland waterways \cite{binnen} to determine the give-way or stand-on priorities.
The following rules are considered:
\bitem
\item Head-on situation: If two vessels are approaching each other on opposite courses in such a way that there is a risk of collision, the vessel not following the starboard side of the fairway shall give-way to the vessel following the starboard side of the fairway. If neither vessel follows the starboard side of the fairway, each shall give-way to vessels on the starboard side so that they pass each other port to port.
\item Crossing situation: If the courses of two ships cross each other in such a way that there is a risk of collision, the vessel not following the starboard side of the fairway shall give-way to the vessel following the starboard side of the fairway. In case none of the ships follows the starboard side of the fairway, the ship approaching from the port side gives way to the vessel approaching from starboard.
\eitem

The details of the priority assignment algorithm on ship $i$ are presented in Algorithm \ref{alg:protocol}.
Firstly, ship $i$ assigns to the ships the priority value 1 if the ship sails on the starboard side of the waterway (steps \ref{alg2:sb-assign1} -- \ref{alg2:sb-assign2}).
If the time to approach the intersection of ship $i$ and ship $j$, are both less than a predefined value $T_{IC}$, then the CROSSING situation is established (step \ref{alg2:crosscon}).
The time to approach the intersection of ship $i$, $T^{IC}_i$, is defined as the time (from present) until ship $i$ reaches a predefined point at the center of the intersection.
We assume that all ships know this center point beforehand.
In CROSSING situation, we increase or decrease the priority value of ship $j$ by $\varpi$ ($\varpi>0$) if ship $i$ comes, respectively, from STARBOARD or PORT side of ship $j$ (steps \ref{alg2:sb-port1} - \ref{alg2:sb-port2}).
It is important that $\varpi<1$, so it does not affect the STARBOARD priority status of ship $j$ and $i$ that was defined in steps \ref{alg2:sb-assign1} -- \ref{alg2:sb-assign2}.
%
%
%
The HEADON status is defined as:
\begin{align*}
	\text{HEADON} =  &\left(\frac{v_{i,n}\cdot v_{j,n}}{\abs{v_{i,n}}\abs{v_{j,n}}}< -\cos(22.5^\circ)\right) \wedge (\text{TCPA}\leq T_{ho})\\
    &\wedge(\text{ship $j,i$ are in the same waterway}) ,
\end{align*}
with TCPA is the time to closest point of approach between two ships, and $T_{ho}$ being a predefined positive scalar.
Here, a ship $j$ is in the same waterway with ship $i$ if $y_{\min} \leq y_{j,i} \leq y_{\max}$, where $y_{\min}$ and $y_{\max}$ define the boundaries width of the waterway in frame $\{w_i\}$ (see Fig. \ref{coordinate}).
The output of Algorithm \ref{alg:protocol} is the priority list $p\_list_i$.
If $p\_list_i(j)=1$, ship $i$ gives way to ship $j$; otherwise ship $i$ does not have to give way to ship $j$.

It is worth noting that the priority value assigned by Algorithm \ref{alg:protocol} is based on traffic rules. 
Therefore, it could be modified based on the traffic rules that are applied.
Additionally, Algorithm \ref{alg:protocol} must satisfy the condition of consistency.
That is, the priority of ship $i$ over ship $j$ that is calculated on ship $i$ must consistent with that is calculated on ship $j$.
This condition means that if Algorithm \ref{alg:protocol} on ship $i$ gives that if ship $i$ gives way to ship $j$, then Algorithm \ref{alg:protocol} on ship $j$ must be ship $j$ stand-on over ship $i$.
In other words, the condition is equivalent to ``if $\rho_{j,i} > \rho_{i,i}$ then $\rho_{j,j} > \rho_{j,i}$''.
In rare cases, if this condition is not satisfied, it could lead to a deadlock situation, which is addressed next, in Section \ref{sec:comm-deadlock}.
\subsection{Resolving deadlock situations}\label{sec:comm-deadlock}
\begin{figure}[!t]
	\centering
	\includegraphics[width=0.8\linewidth]{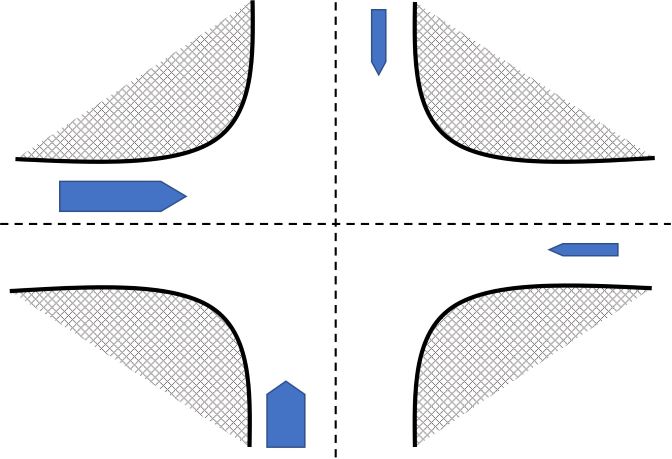}
	\caption{Example of deadlock situation: Each ship wait for the ship comes from their starboard side.}
	\label{deadlock}
\end{figure}

From the step 5 of the TAPD protocol, a C-CAS's iteration starts from ships with the lowest priority.
However, there are rare situations in which a ship with the lowest priority does not exist.
For example, in Fig. \ref{deadlock}, each ship gives way to the ship that comes from the starboard side; none of the four ships will make the first decision.
We call this a deadlock situation.
This situation can be detected with a coordinator that stores and compares the priority points of all ships.
However, our protocol is developed in a distributed manner without a coordinator.
Therefore, we detect and resolve the deadlock situation through the following steps::
\begin{enumerate}
	\item A timer is started when a ship waits for other lower-priority ship decisions. The deadlock is detected if, after a specific time, $T_{DL}$, none of the ships within the network updates its decision.
	\item When a deadlock occurs, each ship exchange the information of its weight and compare with neighboring ships. At this time, the ship with the lightest weight and did not make decision becomes the lowest-priority ship and makes the first decision.
\end{enumerate}

In this process, we use the weight of the ship as the secondary criteria to determine the priority between ships since it is stated in \cite{binnen} that small ships must give way to bigger ships.
In practice, other criteria can be used as a secondary criteria, e.g, length, beam, speed or draft of ships.
\section{Collaborative collision avoidance algorithm}\label{sec:colab}
In the second layer of our framework a distributed MPC-based collaborative collision avoidance algorithm for autonomous ships in inland waterway is embedded.
We use the control scheme presented in Fig. \ref{CAS-protocol}, including line-of-sight (LOS) guidance \cite{fossen11}, CAS, and autopilot.
In this scheme, LOS guidance keeps the ship on track with the predefined waypoints, while CAS helps the ship to avoid collisions.
%
Unlike the MPC-based CASs of \cite{chen18,ferranti_22}, where CASs take responsibility for guiding a ship from waypoint to waypoint, in our approach this task is handled by LOS guidance.
In other words, the only task of the CAS of ship $i$ is modifying the trajectory to guarantee a collision-free path for the that system.
This task is achieved by adding a cross-track offset ($u^y$) and speed modification ($u^s$) to the LOS system so that the OS will avoid an obstacle by sailing parallel with the guiding line.
%


\subsection{Kinematic model of ship $i$}
First let us define the kinematic model of ship $i$ with respect to the path coordinate frame $\{w_i\}$ based on the setpoint filter model \cite{lutz_10} as follows:
\begin{align}
	\begin{split}
		\xii{x}(k) &= \xii{x}(k-1) + \Delta T u_i^s(k)U_i^d\cos(\xii{\chi}(k)),\\
		\xii{y}(k) &= \xii{y}(k-1) + \Delta T u_i^s(k)U_i^d\sin(\xii{\chi}(k)),\\
		\xii{\chi}(k) &= \xii{\chi}(k-1) + \frac{\Delta T}{T_{1}} \cdot \\
        & \Big[\chi^{max}_i\tanh(K_e (u_i^y(k) - \xii{y}(k-1)))- \xii{\chi}(k-1)\Big],
	\end{split}
	\label{kinematic_model}
\end{align}
where $U_i^d, \chi^{max}_i$ are, respectively, the nominal surge speed of ship $i$ and the maximum steering angle that ship $i$ can achieve in a sampling period $\Delta T$, and $T_1, K_e$ are positive constants depending on the hydrodynamics of the vessel and the controller tuning. 
Furthermore, $u_i(k) = [u_i^y(k), u_i^s(k))]^\top$ denotes the control action of ship $i$, where $u_i^y$, $u_i^s$ are the cross-track offset and speed modification, respectively.
The kinematic model \eqref{kinematic_model} of ship $i$ can then be rewritten in compact form as:
\begin{align}\label{model_compact}
	\pii{i}(k+1) = f_i(\pii{i}(k),u_i(k)).
\end{align}
%
\subsection{Risk evaluation}
In order to minimize the risk of collision for ship $i$ with neighboring ships, we introduce the collision risk function of ship $i$ with respect to ship $j$, $R_{ij}(\cdot)$.
The value of the risk function rises sharply when the distance between two ships is closer than a predefined value and it is approximately zero otherwise.
This approach is similar to what is used in the SB-MPC algorithm \cite{johansen16} but with a modification to remove the singular point that can cause numerical issues in a gradient-based MPC.
The predicted risk of collision of ship $i$ with respect to ship $j$ at the time step $t_0+k$ from the present time $t_0$ is defined as:
\begin{align}\label{risk_fnc}
	R_{ij}(t_0+k) = \frac{K_{ca}}{\sqrt{1+K_d k}} \exp\left(-\frac{\left(\delta x_{ij}(k)\right)^2}{\alpha_{xj}} - \frac{\left(\delta y_{ij}(k)\right)^2}{\alpha_{yj}}\right),
\end{align}
where $\delta x_{ij}(k) = \ix{x}{i}(t_0+k) - \ix{x}{j}(t_0+k)$, $\delta y_{ij}(k) = \ix{y}{i}(t_0+k) - \ix{y}{j}(t_0+k)$ are the predicted distances between ship $i$ and $j$, respectively, and $K_{ca}$ is predefined constant based on safety criteria that depend on the traffic situation.
Moreover, $\alpha_{xj}$ and $\alpha_{yj}$ are parameters that are associated with the size, shape, and current speed of ship $j$.
Accordingly, the greater the size or the speed of ship, the large the value of $\alpha_{xj}$ and $\alpha_{yj}$.
In the simulations presented in Section \ref{sec:sim}, we choose $\alpha_{xj}$ and $\alpha_{yj}$ greater than, or equal to, the occupied area of ship $j$. 
Increasing these parameters results in increasing the safety domain of ship $i$, but could lead to unnecessary course or speed change, and less efficient use of the waterway's capacity.
Besides, prediction of risk further away from the present time tends to be less accurate.
Therefore, we implement a discount factor $\frac{1}{\sqrt{1+K_d k}}$, with $K_d \geq 0$, to reduce the weight of the collision risk in the our MPC formulation as $k$ increases.
It is worth mentioning that, in most cases, we do not have $R_{ij} = R_{ji}$, because $\alpha_{xj} \ne \alpha_{xi}$ or $\alpha_{yj} \ne \alpha_{yi}$.
\subsection{Problem formulation}\label{sec:colab-probform}
%

\begin{figure}[!t]
		\centering
		\subfloat[$\Bc_1(x,r)$ with $r=1$]
		{\centering
			\includegraphics[width=0.8\linewidth]{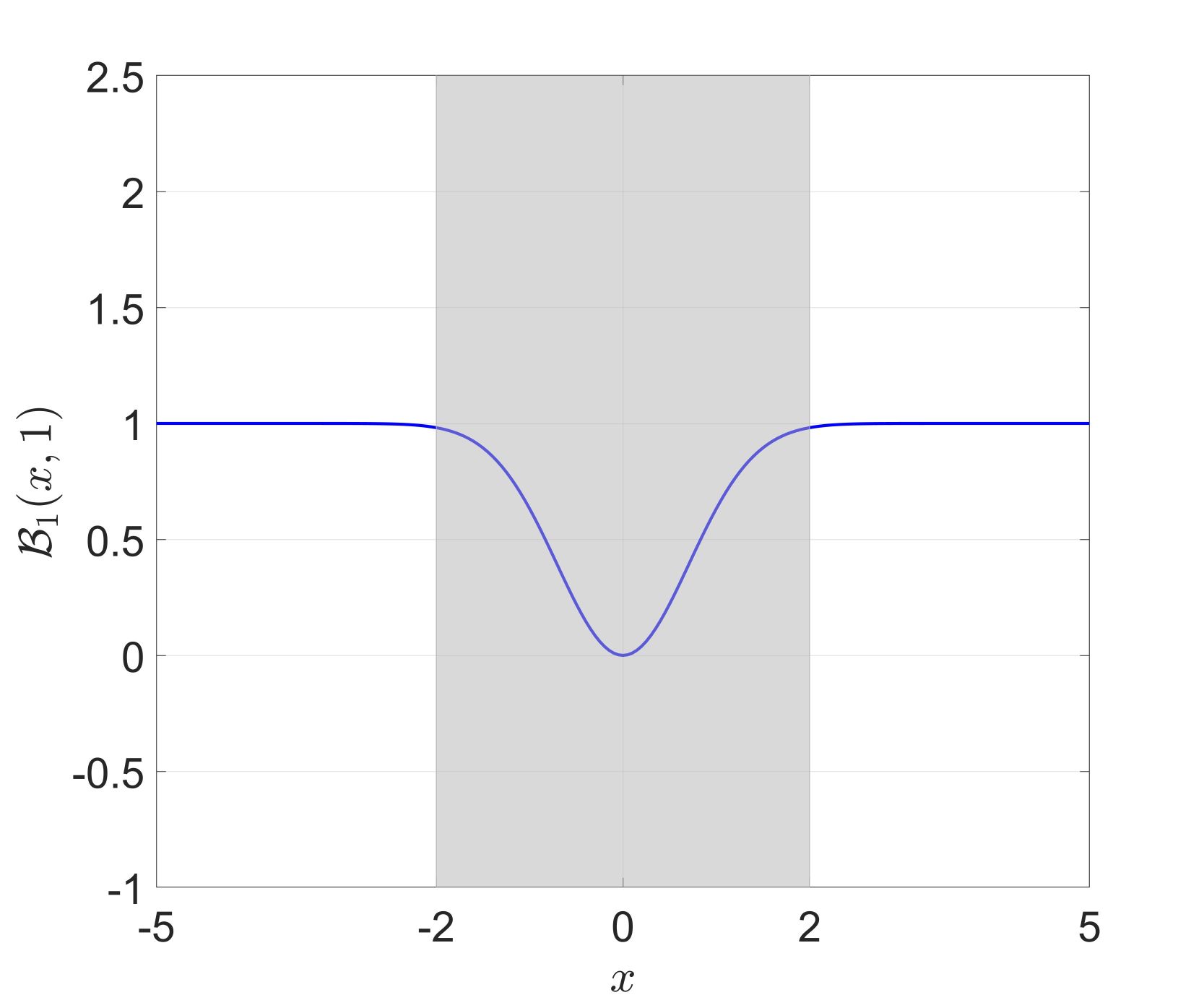}
			\label{beta1}}
		\hfil
		\subfloat[$\Bc_2(x,r_1,r_2)$ with $r_1=0.1$, $r_2=2-\pi$.]
		{\centering
			\includegraphics[width=0.8\linewidth]{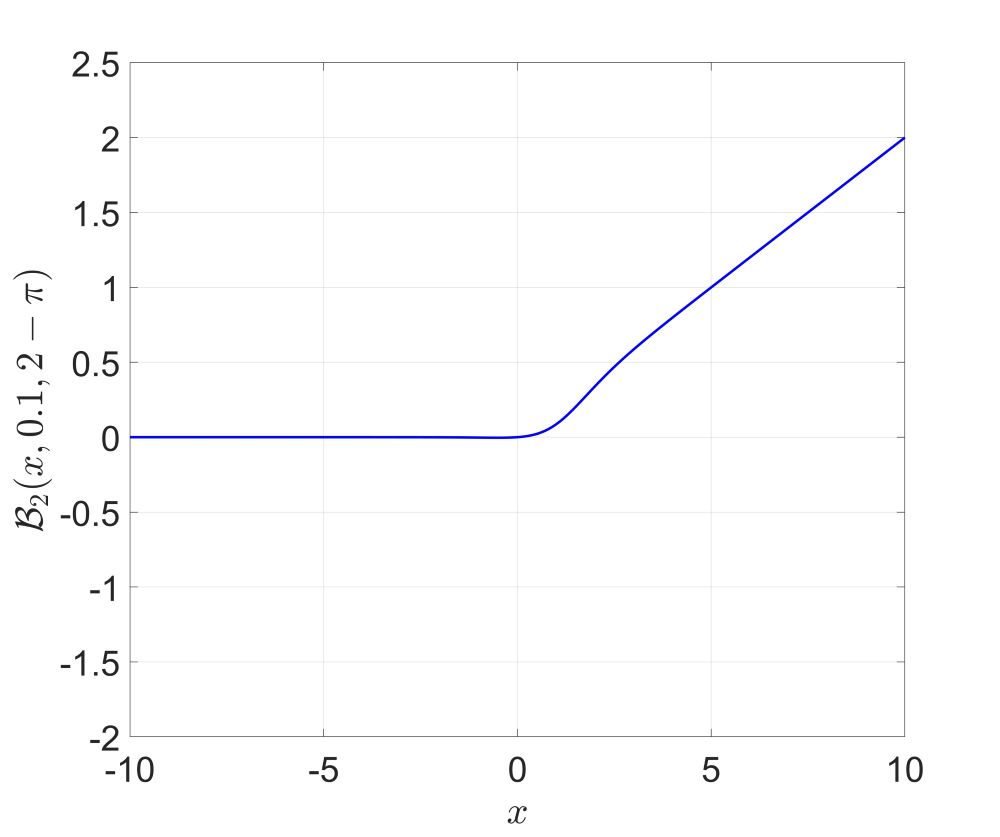}
			\label{tanh}}
		\hfil
		\caption{Behavior functions $\Bc_1(x,r)$ and $\Bc_2(x,r_1,r_2)$.}
	\end{figure}
The main task of the CAS is to avoid collision with consistent behaviors that comply with traffic rules.
Thus, we propose the following cost function of ship $i$ for the MPC formulation:
\begin{align}
	\Jc_i(\bpi{i},\tilde{u}_i) =\Jc_i^{ca}(\bpi{i}) + \Jc_i^{e}({\tilde{u}}_{i}) + \Jc_i^{b}({\tilde{u}}_{i}),
	\label{cost}
\end{align}
where $\bpi{i}(t)$ and $\tilde{u}_i(t)$ are vectors containing the system state and input over the control horizon, i.e., $\tilde{u}_i(t) =[u_i^\top(t), u_i^\top(t+1),...,u_i^\top(t+N-1)]^\top$, $\bpi{i}(t) = [\pii{i}^\top(t), \pii{i}^\top(t+1),...,\pii{i}^\top(t+N)]^\top$.
Moreover, $\Jc_i^{ca}(\bpi{i})$ is the sum of collision risks with respect to all neighboring ships over the horizon, i.e., $\Jc_i^{ca}(\bpi{i}) = \sum_{k=1}^{N+1}  \left(\sum_{j \in \Mc \backslash \{i\}} R_{ij}(t+k)\right)$.
It should be mentioned that $\Jc_i^{ca}(\bpi{i})$ also depends on the state of the neighboring ships.
However, in the cost function of ship $i$, the state of ship $j$ is not a decision variable and is considered as a given parameter instead.
$\Jc_i^{e}({\tilde{u}}_{i})$ represents the cost of collision avoidance control actions, necessary for reflecting that ship $i$ should only change the course or speed when it can significantly decrease the collision risk.
%
$\Jc_i^{e}({\tilde{u}}_{i})$ is defined as follows:
\begin{align}
	 \Jc_i^{e}({\tilde{u}}_{i}) = \sum_{k=1}^{N+1} \Big[&K_y\left(u^y_i(t+k) - u^y_i(t+k-1)\right)^2  \non\\
     &+ K_s\left(1-u^s_i(t+k)\right)^2 \Big],
\end{align}
with $K_y$, $K_s>0$ being control parameters.
On the one hand, smaller values of $K_y$ and $K_s$ make the ship overreact with collision risk, i.e., making an unnecessary action.
On the other hand, larger values reduce the magnitude of course/speed change of the ship when facing a risk of collision.
We can prioritize one collision avoidance action over another by changing the ratio $\frac{K_y}{K_s}$.
For example, ship $i$ will prioritize to change course rather than changing speed if $\frac{K_y}{K_s}$ is reduced and vice versa.

In order to make the behavior of ship $i$ adequately represent traffic rules and be more visible to neighboring ships, i.e., the change in course or speed should be of a sufficiently large magnitude instead of a sequence of small changes; we define $\Jc_i^{b}({\tilde{u}}_{i})$ as follows:
\begin{align}
	\Jc_i^{b}({\tilde{u}}_{i}) = \sum_{k=1}^{N+1} &\Big[ \mu_1 \Bc_1\left(1-u^s_i(t+k),\gamma_s\right) \non\\
    &+ \mu_2 \Bc_1\left(u^y_i(t+k) - u^y_i(t+k-1),\gamma_y\right)  \non\\
	& + \Bc_2\left(u^y_i(t+k-1) - u^y_i(t+k),\mu_3\right)\Big]. \label{behav}
\end{align}
The first and second terms in \eqref{behav} represent that if the collision avoidance decision is made, it must be large enough to be observable by other neighboring ships.
$\Bc_1(x,r)$ is defined as $\Bc_1(x,r) = 1- \exp\left(\frac{-x^2}{r}\right)$, with $r>0$ is a parameter.
As illustrated in Fig. \ref{beta1}, if the solution lies within the grey area, it can be easily be attracted toward the minimum point at zero.
However, if the solution lies outside the grey area, the solution now lies on a local minimum point (as the gradient at this area is approximately zero).
Hence, $\Bc_1(x,r)$ penalizes only the solution with small magnitude, but not those with large enough magnitude.
%
%
The third term penalizes the port side steering behavior of ship $i$.
The behavior function $\Bc_2(x,r_1,r_2)$ is defined as $\Bc_2(x,r_1,r_2) = r_1 x[\tanh(x+r_2)+1]$, where $r_1,r_2$ are constant parameters.
As shown in Fig. \ref{tanh}, the value of the function will increase as the ship steers toward the port side, i.e., $u^y_i(t+k-1) >u^y_i(t+k)$.
%
%
Moreover, because $\Bc_2(x,r_1,r_2)$ increases slowly, ship $i$ can still steer port if it can significantly reduce risk.

We formulate the distributed MPC collision avoidance problem of ship $i$, $i \in \Mc$, with cost function \eqref{cost} as follows:
\begin{subequations}\label{prob1}
\begin{align}
		\min_{\bpi{i},{\tilde{u}}_i} \quad & \quad \Jc_i(\tilde{p}_{i,i},\tilde{u}_i) \\
	\textrm{s.t.:} \quad  &  \quad  \pii{i}(t+k+1) = f_i(\pii{i}(t+k),u_i(t+k)),  \label{con_p1}\\
				   \quad  &  \quad  \pii{i}(t) = p_{i,i}^{init},  \label{con_p2}\\
				   \quad	 &  \quad  {u}_i(t+k) \in U_i, \label{con_u} \\
	 			   \quad  &  \quad \bpi{i} = \tilde{R}_i (\xi_{i} - \bar{\eta}_{i}^{WP^-} ),  \label{con1a}
\end{align}
\end{subequations}
where $p_{i,i}^{init}$ is the current position of ship $i$, $U_i$ defines the boundary set for control inputs ${u}_i(t+k)$; $\xi_{i}$ is the global variable that is sent to other ships. It represents the predicted position of ship $i$ with respect to $\{n\}$ over the prediction horizon, and is defined as $\xi_{i} = \tilde{R}_i^{-1}\bpi{i} + \bar{\eta}_{i}^{WP^-}$.
Moreover, $\bar{\eta}_{i}^{WP^-} = \bos{1}_{N+1} \otimes {\eta}_{i}^{WP^-}$.
Besides, $\tilde{R}_i$ is a block diagonal matrix with each block being the rotation matrix $R_i$:
\begin{align*}
	\tilde{R}_i = \bbm R_i & &0 \\
					    &\ddots&\\
					    0 & & R_i\ebm \in \Rbb^{3(N+1)\times 3(N+1)}.
\end{align*}
%
%
%
%
Let us introduce $\btp{i}$ as the global variable that stores the position of ship $i$ with respect to $\{w_i\}$, i.e., $\btp{i} = \tilde{R}_i (\xi_{i} - \bar{\eta}_{i}^{WP^-})$.
%
%
Problem \eqref{prob1} is now rewritten in compact form as follows:
\begin{subequations} \label{prob2}
	\begin{align}
		\min_{\tilde{u}_i,\bpi{i}} \quad & \quad  \Jc_i(\tilde{p}_{i,i} ,\tilde{u}_i)\\
		\textrm{s.t.:}  \quad    &  \quad  [\tilde{u}_i^\top,\bpi{i}^\top]^\top \in \Gbf_i \label{cons_all} \\
						\quad    &  \quad  \bpi{i}  = \btp{i}, \label{11c}
	\end{align}
\end{subequations}
where (\ref{11c}) is equivalent to (\ref{con1a}), and
\begin{align}
\Gbf_i:=\left\{[\tilde{u}_i^\top,\bpi{i}^\top]^\top|\text{\eqref{con_p1}, \eqref{con_p2}, \eqref{con_u} are satisfied}\right\}
\end{align}
is the feasible region for ship $i$.
%
%
%
%
%
Then, we have the NADMM update for the controller of ship $i$ at iteration index $s$:
\begin{subequations} \label{ADMM-update}
\begin{align}
	z^{s+1/2}_{i} &= z^{s}_{i} - \beta(1-\lambda)\left(\bpi{i}^{s} - \btp{i}^{s}\right) \label{update_z1}\\
	\bbm \tilde{u}_i^{s+1} \\ \bpi{i}^{s+1}\ebm &= \text{argmin}_{ [\tilde{u}_i,\bpi{i}]^\top \in \Gbf_i} \Big\{\Jc_i(\bpi{i},\tilde{u}_i) \non \\
	&+\left<z_i^{s+1/2},\left(\bpi{i} - \btp{i}^{s}\right)\right> + \frac{\beta}{2}\lVert\bpi{i} - \btp{i}^{s}\rVert^2\Big\} \label{update_local}\\
	z^{s+1}_{i} &= z^{s+1/2}_{i} + \beta\left(\bpi{i}^{s+1} - \btp{i}^{s}\right), \label{update_z2}\\
	\btp{i}^{s+1} &=\bpi{i}^{s+1} + \frac{1}{\beta}z^{s+1}_{i}. \label{update_global}
\end{align}
\end{subequations}
It should be noted that when performing the NADMM update, each agent only update their decision variable and does not change others' decisions.
Therefore, only parts of $w$ and $v$ are updated (in \eqref{update_local} and \eqref{update_global}, respectively) each time an agent performs \eqref{ADMM-update}.

The details of the collaborative collision avoidance algorithm are given in Algorithm \ref{alg1}.
The algorithm is executed after the communication protocol has established the priority list.
%
%
According to the priority list ship $i$ has to wait other ships with lower priority.
At each iteration $s$, if all lower priority ships have made their decision and the number of iterations has not reached the maximum (step \ref{alg1:checkwait}), then ship $i$ can execute the algorithm (step \ref{alg1:updateADMM}).
Before performing an update, ship $i$ needs to know the future position of ship $j$, i.e., $\bpi{j}$, to calculate the risk function $R_{ij}(\cdot)$ as in \eqref{risk_fnc}.
The future position of ship $j$ is obtained via communication if it is available (step \ref{alg1:update_comm}) or from a motion prediction algorithm if it is not available.
Different prediction algorithms can be used to predict future motion of neighboring ships based on information from sensors and AIS system (see \cite{tran_collision_2023}).
Here, we adopt a constant velocity model to predict the motion of neighboring ships.
In step \ref{alg1:update_predict}, the prediction algorithm is called.
%
%
The decision of ship $i$ is then transformed into the inertial frame (step \ref{alg1:transform}) and broadcast to neighboring ships (step \ref{alg1:transmit}).
%
%
\begin{algorithm}[!t]
	\caption{Collaborative collision avoidance} \label{alg1}
	\hspace*{\algorithmicindent} \textbf{Input:} For $i,j \in \Mc, i \neq j$, initialize $\textbf{p}^{1}_{ii},z^{1}_{ij}$ based on trajectory prediction. 
	\begin{algorithmic}[1]
		\STATE $iter(i) := 0 ~ \forall i \in \Mc$.
		\FOR{s = 1,2,...}
			\FORALL{$i \in \Mc$}
				\IF{$wait(i) == \FALSE$ \AND $ADMM_{DONE}(i) == \FALSE$} \label{alg1:checkwait}
					\FORALL{$j \ne i$}
					\IF{$\xi_{j}$ is available}
						\STATE $\bpi{j} := \tilde{R}_i (\xi_{j} - \bar{\eta}_{i}^{WP^-})$ \label{alg1:update_comm}
					\ELSE
						\STATE $\bpi{j} := trajectory\_prediction(\pii{j}(t))$ \label{alg1:update_predict}
					\ENDIF
					\ENDFOR
					\STATE update according to \eqref{ADMM-update} \label{alg1:updateADMM}
					\STATE $\xi_{i} := \tilde{R}_i^{-1}\bpi{i} + \bar{\eta}_{i}^{WP^-}$ \label{alg1:transform}
					\STATE Transmit data $\xi_{i}$ to all ship $j \in \Mc_i$. \label{alg1:transmit}
					\STATE $iter(i) := iter(i)+1 $.
					\IF{$iter(i) == iter_{max}$}
						\STATE $ADMM_{DONE}(i) = \TRUE$
						\STATE $iter(i) := 0$.
					\ENDIF
				\ELSE 
					\STATE $\tilde{u}_i^{s+1}:=\tilde{u}_i^{s}$
					\STATE $\bpi{i}^{s+1}:=\bpi{i}^{s}$
				\ENDIF
			\ENDFOR
		\IF {$ADMM_{DONE}(i) == \TRUE~ \forall i \in \Mc$}
			\STATE \textbf{break}.
		\ENDIF
		\ENDFOR
	\end{algorithmic}
\end{algorithm}
\subsection{Convergence of Algorithm \ref{alg1}}
We present in Theorem \ref{theorem:converge} the convergence analysis for Algorithm \ref{alg1}.
This proof of convergence guarantees that the solution provided by Algorithm \ref{alg1} helps ships to collaboratively reduce collision risks.
\begin{theorem}\label{theorem:converge}
	Assume Problem \eqref{prob2} has feasible solutions for all $i \in \Mc$. Then the solution provided by Algorithm \ref{alg1} with $\lambda \in (0,2)$, $\beta > 2L$, and $L>0$ converges asymptotically to a (locally) optimal solution.
\end{theorem}
\begin{proof}
	See Appendix \ref{proof}.
\end{proof}

Algorithm \ref{alg1} works when a solution exists to avoid collision.
If such a solution does not exist, e.g., the waterway is too narrow to do any course change, human interference is required.
\section{Simulation experiments} \label{sec:sim}
This section illustrates the performance of the proposed framework through several simulation experiments.
Two typical scenarios are being used to evaluate the proposed framework: head-on and intersection crossing.
In head-on scenarios, by evaluating different control parameters, we show how these parameter affect ships' behavior.
In intersection crossing scenarios, we show how the complexity of the traffic situation could affect the behavior of ships.

%
%
%

%

The local MPC problem \eqref{update_local} is solved using the Casadi toolbox \cite{Andersson2019} with the \textit{ipopt} solver \cite{wachter_implementation_2004}, i.e., an open source interior point optimizer software package for large-scale nonlinear optimization.
The simulations are implemented in Matlab2023a running on a PC with an Intel(R) Core(TM) i7-11850H and 32GB of RAM.

\subsection{Head-on scenarios}\label{HO-scenarios}
\begin{figure}[!t]
	\centering
	\includegraphics[width=0.8\linewidth]{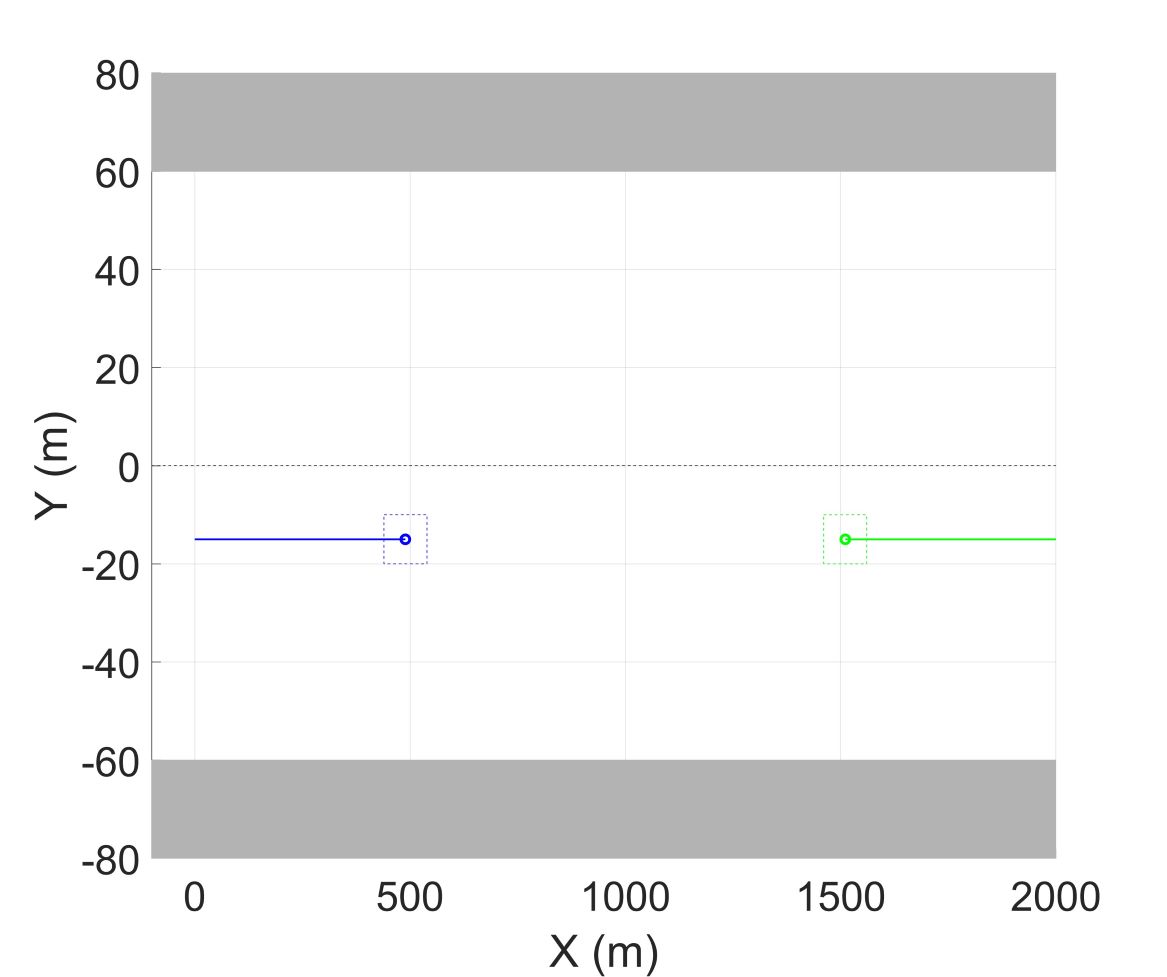}
	\caption{Head on situation in inland waterway between two ships. Ship 1 and 2 are illustrated in blue and green dots, respectively. The rectangle around each ship is the safety area. The waterway is illustrated in white, and the grey area is where a ship cannot sail. The dash line at $Y=0m$ is the central line divide port and starboard side of a waterway.}
	\label{ho}
\end{figure}
In this scenario, two ships sail in the same waterway but in opposite directions (see Fig. \ref{ho}).
This scenario requires each ship to adopt a course change.
According to the traffic rules mentioned in Section \ref{sec:comm-prot}, a ship sailing on the starboard side of the waterway has the right to stand-on.
Consequently, the expected resulting paths should be that ship 2 gives way by steering toward the starboard while ship 1 keeps its way.
Additionally, this scenario considers the waterway to be wide enough for ships to perform a course change.
Otherwise, if the waterway is too narrow and a feasible solution does not exist, a human decision is needed.

\subsubsection{Impact of control parameters on behavior of ships}
\begin{figure*}[!t]
	\centering
	\subfloat[\label{ho1: ky1e-2}]
    {\centering
		\includegraphics[width=0.3\linewidth]{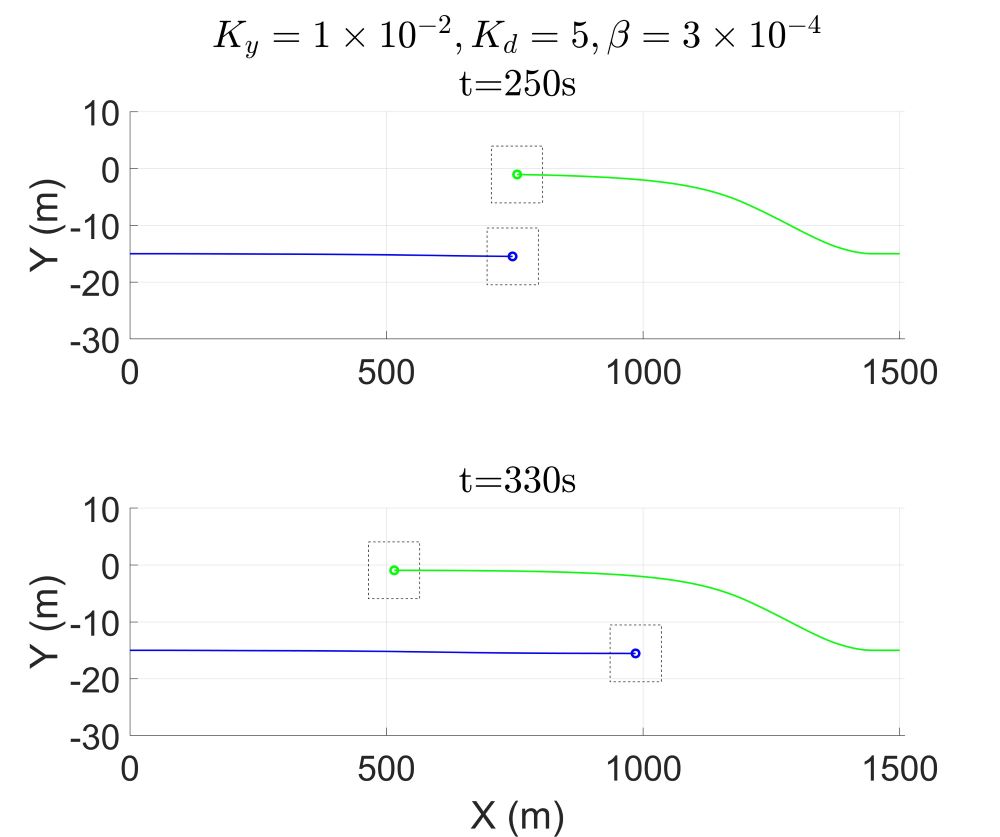}
		}
    \hfill
    \centering
    \subfloat[\label{ho1: ky1e-3}]
	{\centering
		\includegraphics[width=0.3\linewidth]{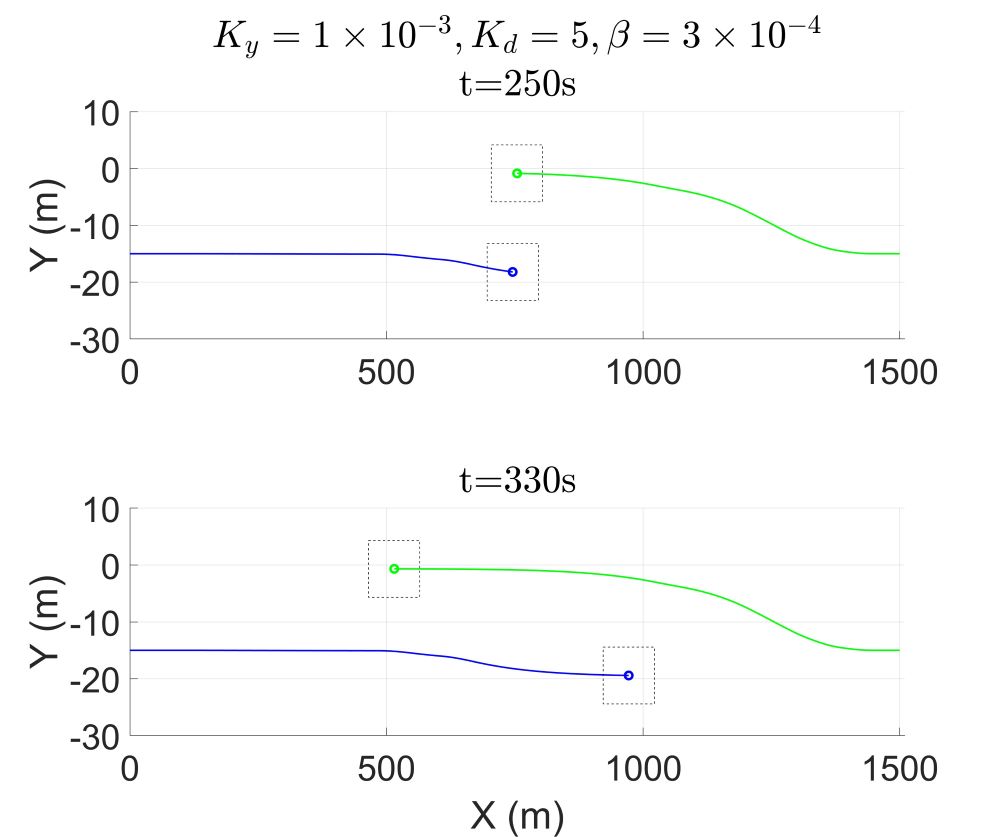}
		}
	\hfill
	\centering
	\subfloat[\label{ho1: ky3e-2}]
		{\centering
		\includegraphics[width=0.3\linewidth]{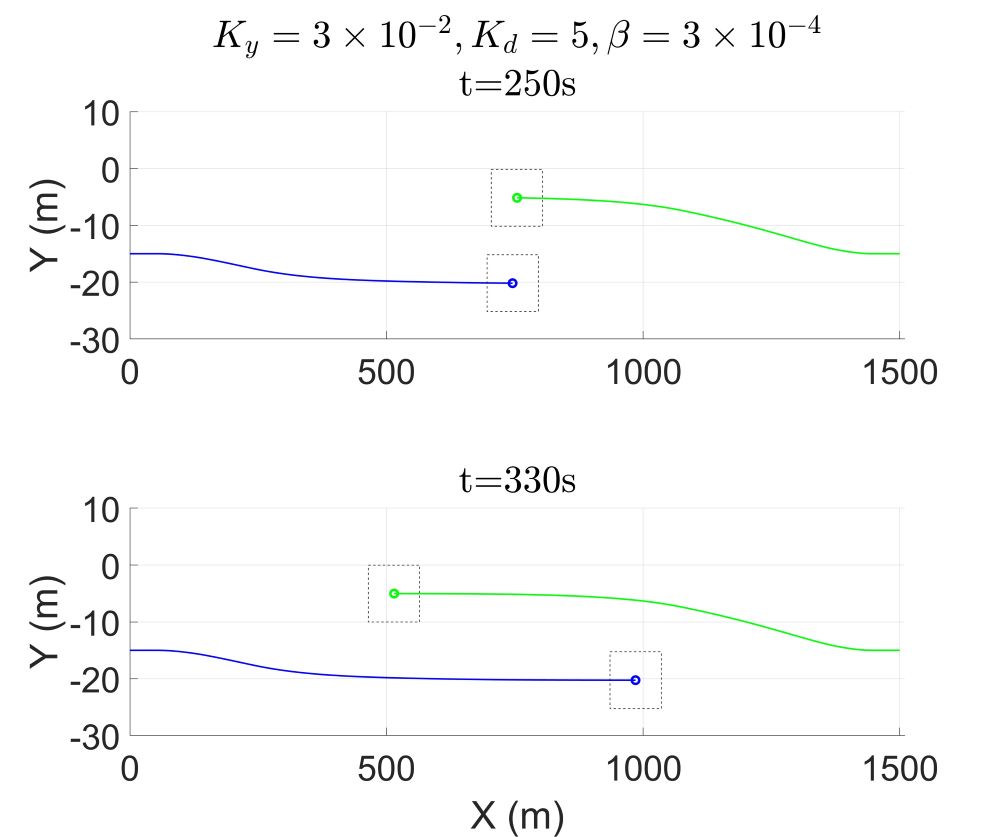}
		}
	\hfill
	\centering
	\subfloat[\label{ho1: ky6e-2}]
		{\centering
		\includegraphics[width=0.3\linewidth]{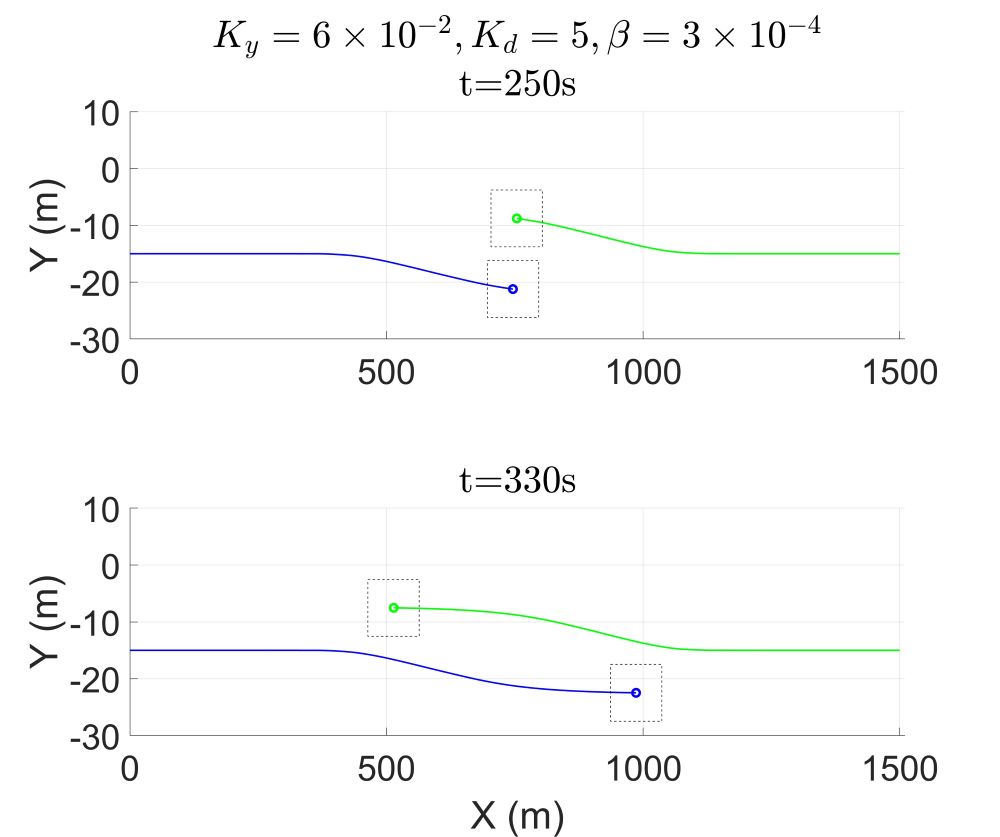}
		}
    \hfill
	\centering
	\subfloat[\label{ho1: kd20}]{
		\centering
		\includegraphics[width=0.3\linewidth]{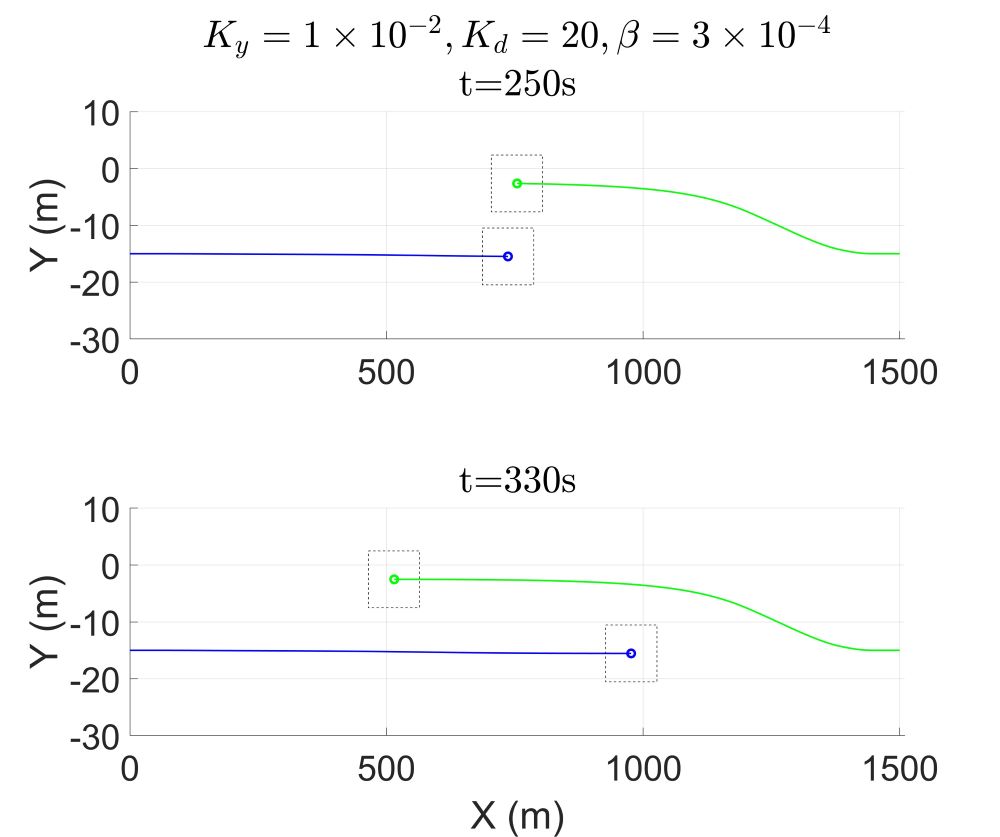}
		}
	\hfill
	\centering
	\subfloat[\label{ho1: kd60}]{
		\centering
		\includegraphics[width=0.3\linewidth]{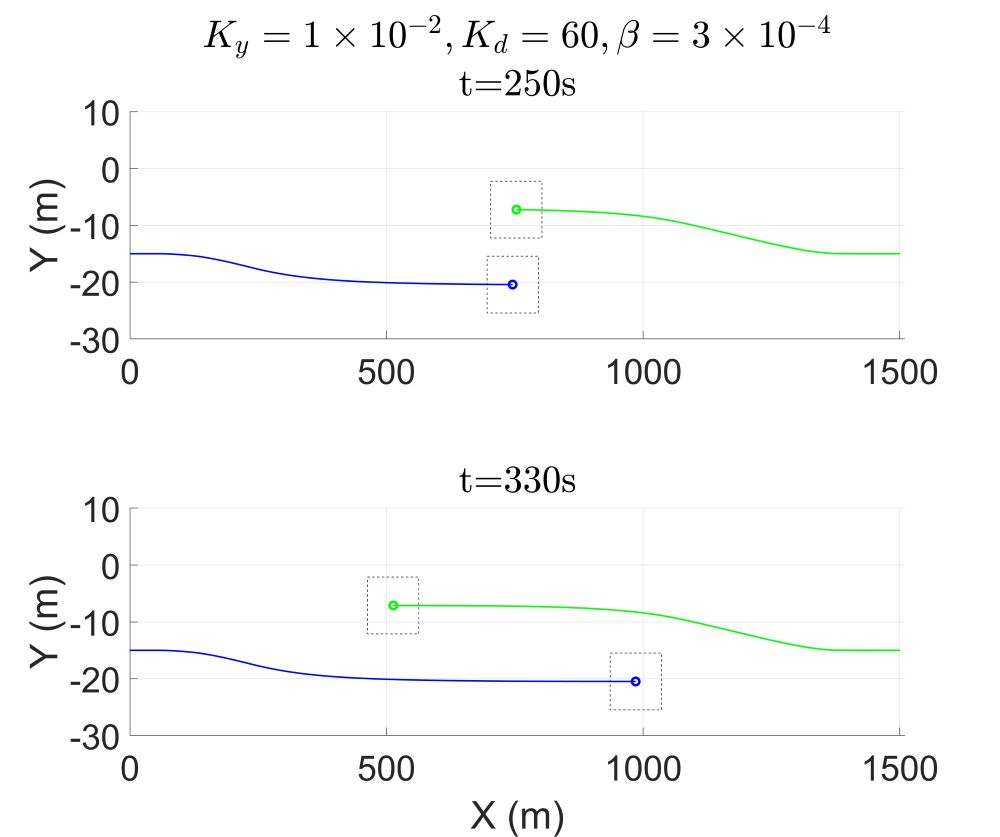}
		}
	\hfill
	\centering
	\subfloat[\label{ho1: beta3e-5}]{
		\centering
		\includegraphics[width=0.3\linewidth]{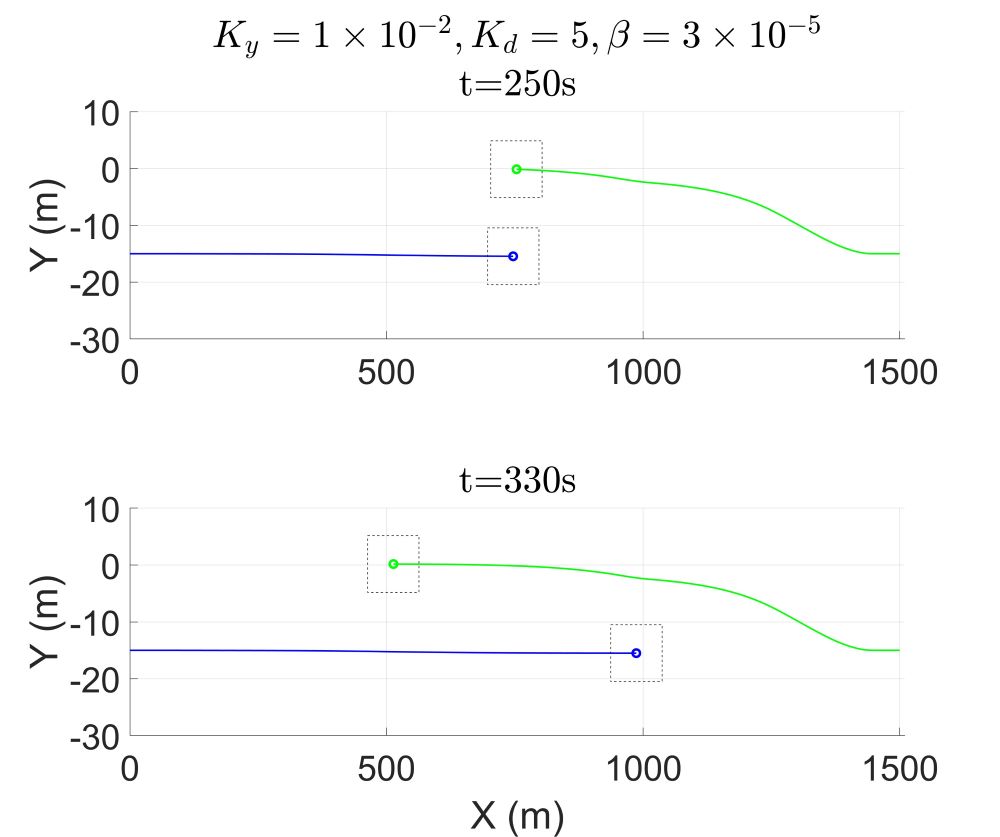}
		}
	\centering
	\subfloat[\label{ho1: beta15e-4}]{
		\centering
		\includegraphics[width=0.3\linewidth]{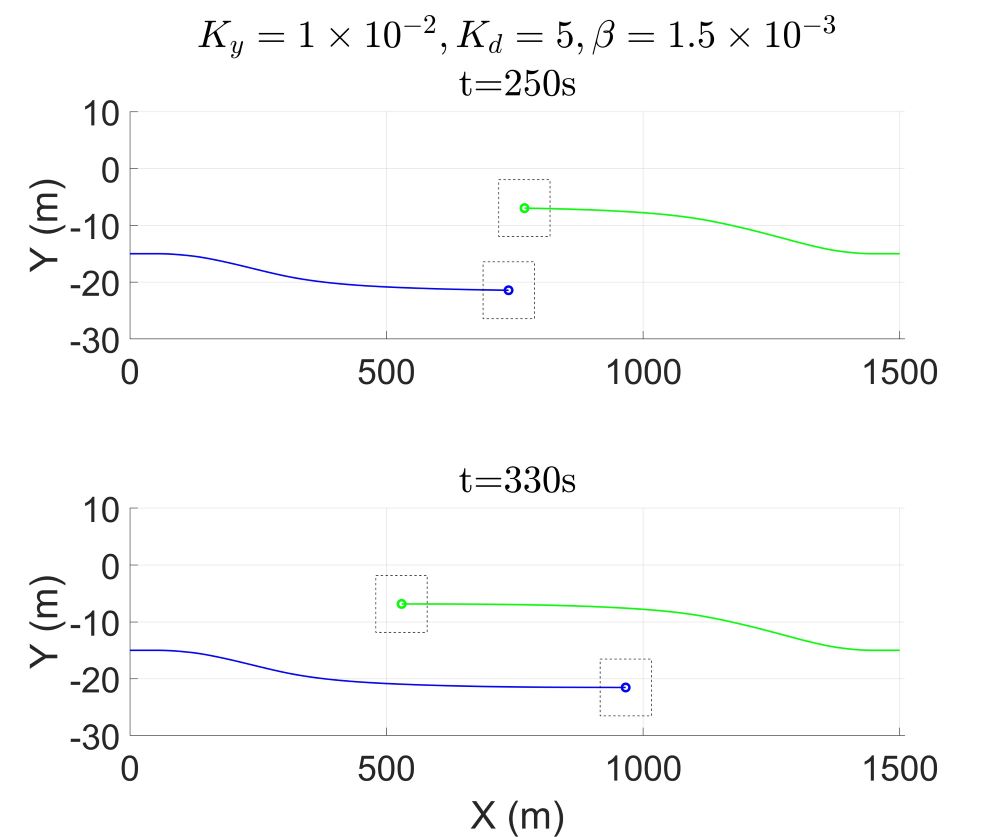}
		}
	\centering
	\subfloat[\label{ho1: beta6e-3}]{
		\centering
		\includegraphics[width=0.3\linewidth]{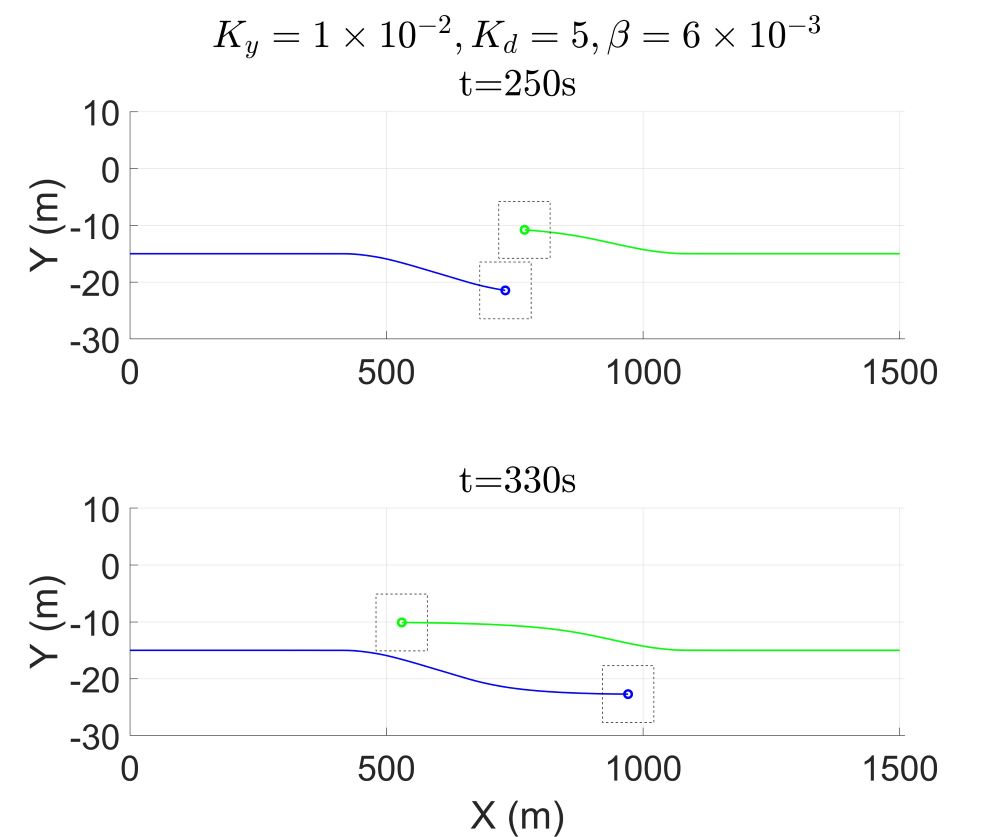}
		}
	\caption{Head-on situation with different sets of parameters. Ship 1 and 2 are illustrated in blue and green dots, respectively.}
	\label{ho1}
\end{figure*}
We investigate the behavior of the proposed collaborative collision avoidance algorithm with different control parameters.
We use the minimum distance in the x-axis, i.e., the distance in the x-axis over whole experiment, and the stand-on and give-way behavior of two ships as evaluation criteria. 
The investigated parameters are $K_y, K_d,$ and $\beta$ as these parameters directly affect the magnitude of the course change of the ship when facing collision risks.
We avoid modifying $K_{ca}, \alpha_{x}, \alpha_{y}$ since they are directly related to the size or shape of neighboring ships, and $K_s = 2\times 10^{-2}$.
Through trial and error, we found the best parameters used as a reference in this scenario are $K_y =10^{-2}$, $K_d = 5$, and $\beta = 3\times 10^{-4}$.
As shown in Fig. \ref{ho1: ky1e-2}, ship 2 takes action early to avoid collision and keep a clear way so that ship 1 does not have to change course.

From Fig. \ref{ho1: ky1e-3}, if we decrease $K_y$ to $10^{-3}$, the stand-on ship tend to overreact with respect to collision risk.
Although ship 2 has made a starboard turn at a safe distance, ship 1 still made an unnecessary course change.
On the other hand, increasing $K_y$ reduces the minimum distance in the x-axis and delays the time of the first action (see Fig. \ref{ho1: ky3e-2} and \ref{ho1: ky6e-2}).
These results also coincide with the analysis in Section \ref{sec:colab-probform}.
Furthermore, the reduction of the minimum distance in the x-axis in collision avoidance action also causes another unwanted behavior: the changing course of the stand-on ship.

While changing $K_y$ could significantly affect a ship's behavior, changing $K_d$ causes a lesser impact.
As shown in Fig. \ref{ho1: kd20}, $K_d$ increasing four times just slightly reduces the minimum distance in the x-axis.
Similarly to $K_y$, a large value of $K_d$ influences the stand-on ship to change course (see Fig. \ref{ho1: kd60}).

Different from $K_y$ and $K_{d}$, a change of $\beta$ does not affect the minimum distance in the x-axis.
However, the give-way and stand-on behavior are greatly affected.
Reducing $\beta$ causes a slightly inconsistent behavior of the give-way ship, while the overall performance is the same (see Fig. \ref{ho1: beta3e-5}).
On the other hand, increasing $\beta$ persuades the stand-on ship to take action (see Fig \ref{ho1: beta15e-4}) or even swaps the role between stand-on and give-way ship (see Fig \ref{ho1: beta6e-3}).

Depending on specific requirements of CAS, one can adjust the aforementioned parameter to achieve the ship's desired behaviors. 
However it should be kept in mind that changing the control parameter could, sometimes, result in violating the traffic rules.
The behavior is still safe, since it is agreed upon among of both ships.
%
\subsubsection{Impact of the number of iterations}
\begin{figure*}[!t]
	\centering
	\subfloat[\label{iter0}]{
		\centering
		\includegraphics[width=0.3\linewidth]{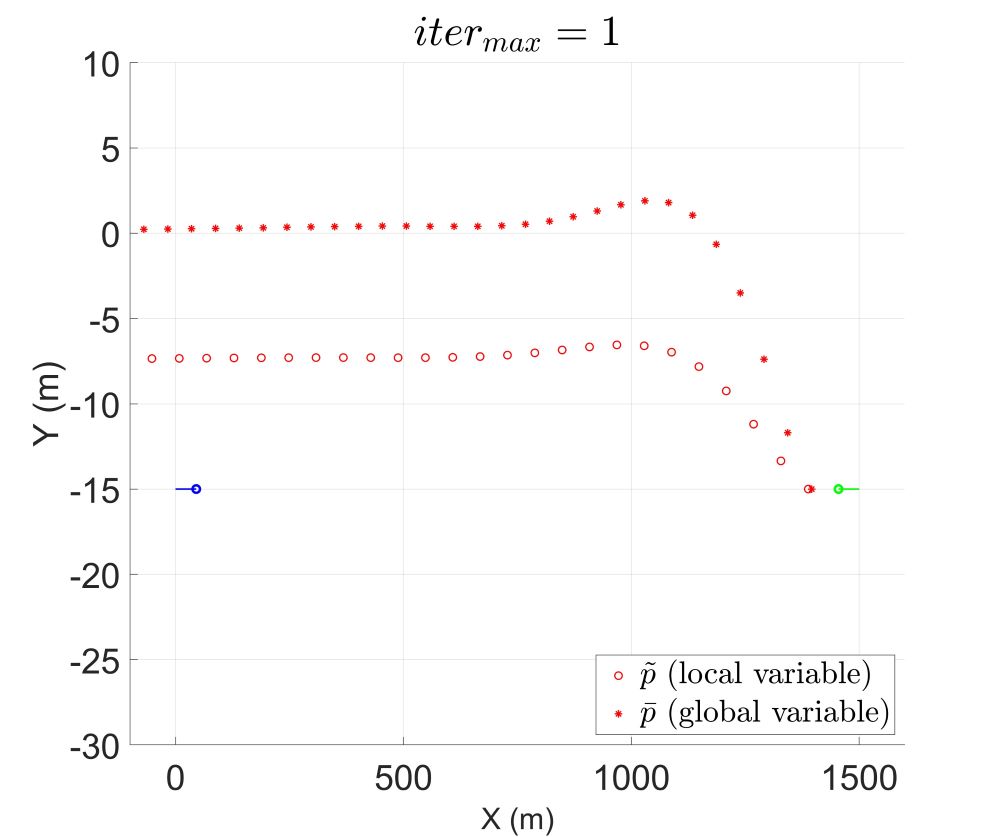}}
	\centering
	\subfloat[\label{iter2}]{
		\centering
		\includegraphics[width=0.3\linewidth]{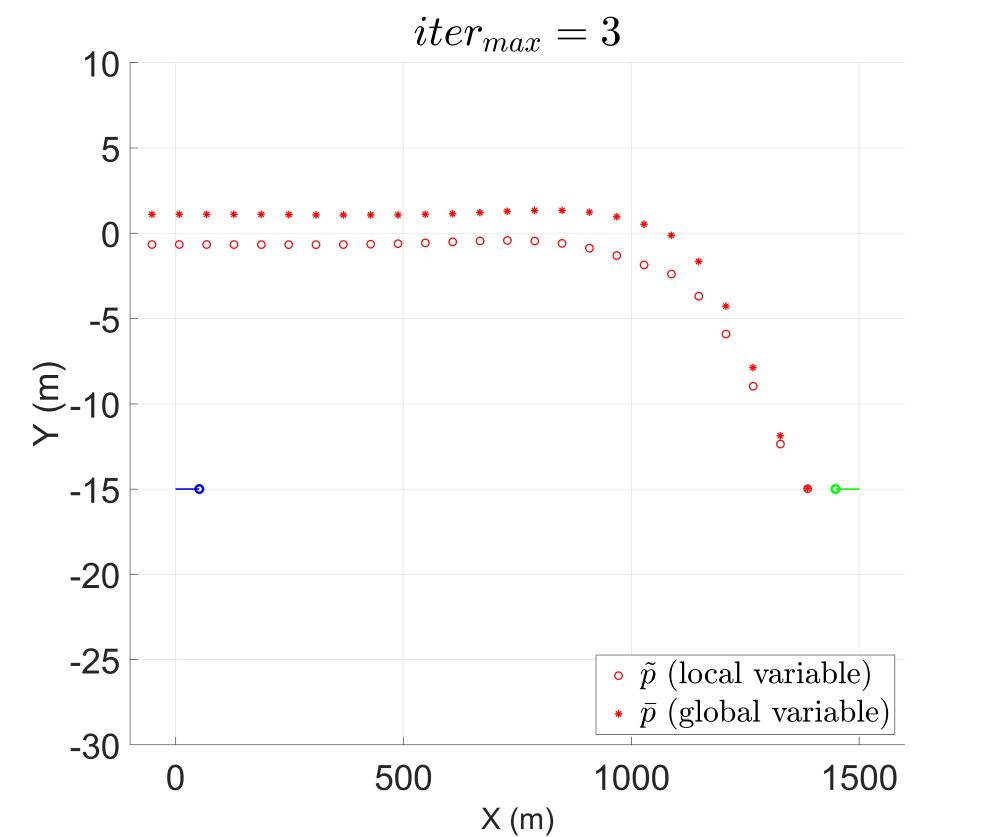}}
	\hfill
	\centering
	\subfloat[\label{iter5}]{
		\centering
		\includegraphics[width=0.3\linewidth]{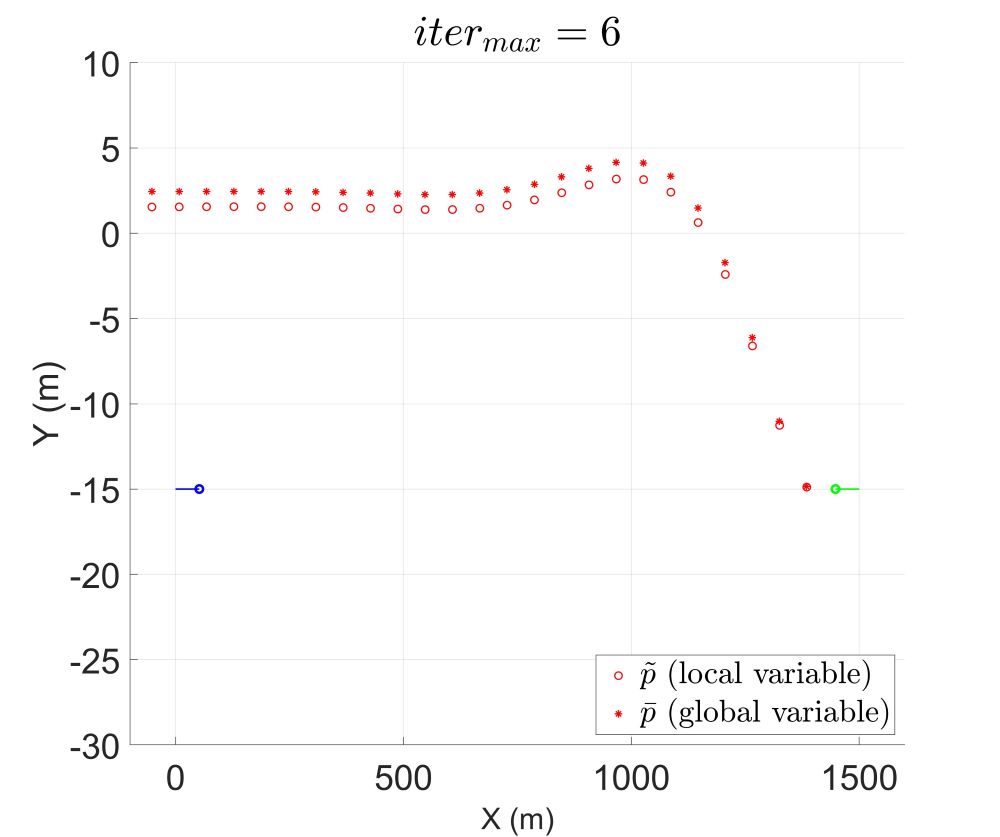}}
	\subfloat[\label{iter10}]{
		\centering
		\includegraphics[width=0.3\linewidth]{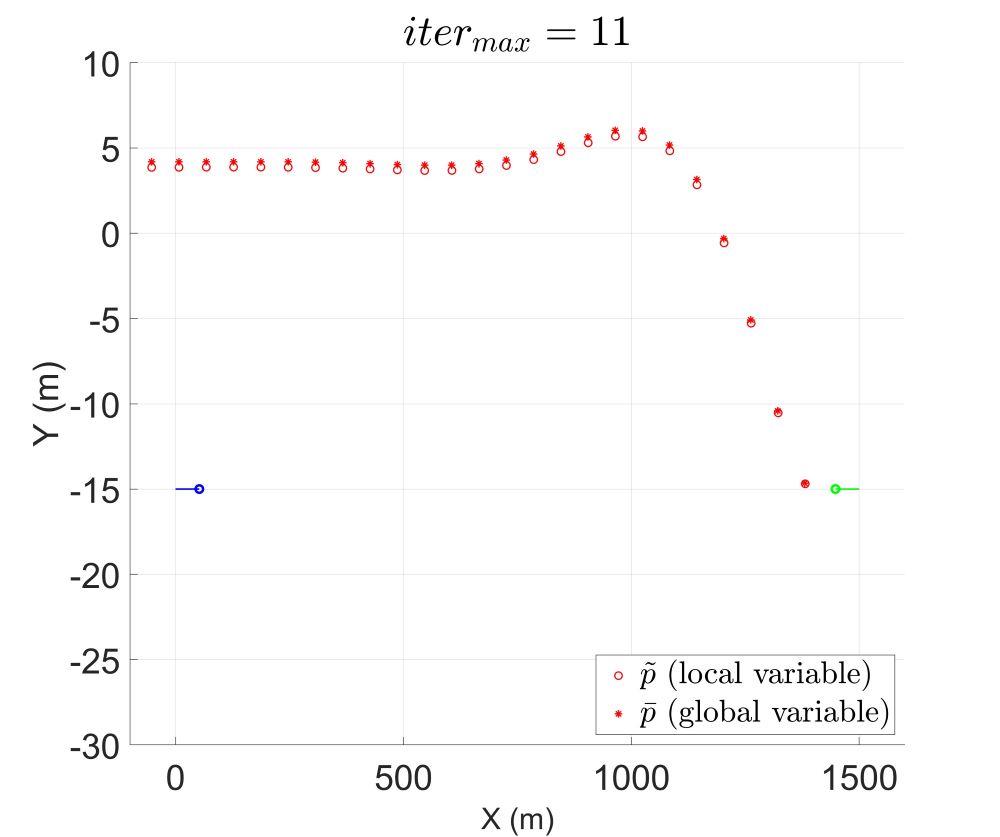}}
	\caption{Difference between $\btp{i}^{s+1}$ and $\bpi{i}^{s+1}$ with different $iter_{max}$. Ship 1 and 2 are illustrated in blue and green dots, respectively.}
	\label{iter}
\end{figure*}

\begin{figure*}[!t]
	\centering
	\subfloat[\label{iter2w}]{
		\centering
		\includegraphics[width=0.3\linewidth]{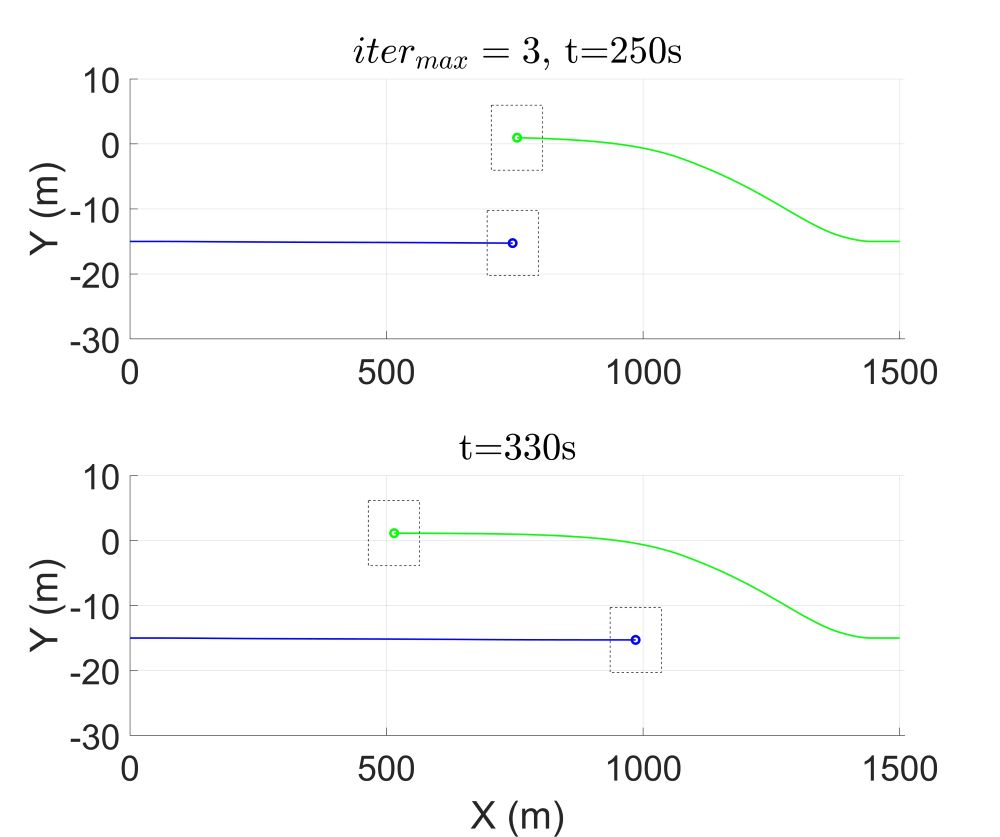}}
	\centering
	\subfloat[\label{iter10w}]{
		\centering
		\includegraphics[width=0.3\linewidth]{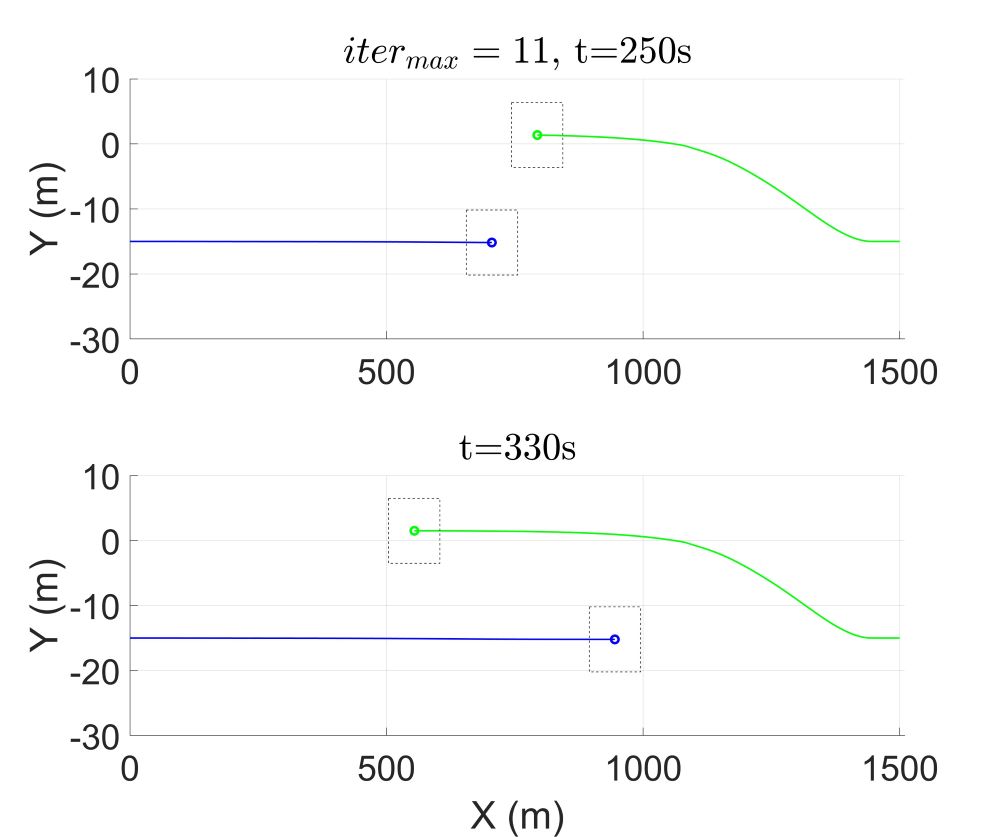}}
	\caption{The solution with different $iter_{max}$. Ship 1 and 2 are illustrated in blue and green dots, respectively.}
	\label{iterw}
\end{figure*}

As we see in equation \eqref{update_global}, the trajectory prediction that is broadcast to neighboring ships ($\btp{i}^{s+1}$), is not equal to the locally predicted trajectory ($\bpi{i}^{s+1}$), unless $z^{s+1}_i = 0$.
The Lagrange multiplier, $z^{s+1}_i$, converges to zero as we increase the maximum number of iterations ($iter_{max}$).
However, when we increase $iter_{max}$, the computation time also increases.
Therefore, to use it in real-time systems, we should balance control performance and computation time.
We evaluated the performance of the proposed algorithm with different $iter_{max}$.
The results are shown in Fig. \ref{iter}.
It is clear that if we perform only one iteration ($iter_{max}=1$), the difference between $\btp{i}^{s+1}$ and $\bpi{i}^{s+1}$ is significant (see Fig. \ref{iter0}).
The difference significantly decreases as $iter_{max}=3$, and when $iter_{max}=11$, the locally predicted trajectory coincides almost completely with the broadcast trajectory prediction.
However, because there is no significant difference between the final solution in case $iter_{max}=3$ compared with $iter_{max}=11$ (see Fig. \ref{iterw}), we may choose $iter_{max}=3$ to reduce the unnecessary computation time.
It is worth mentioning that, in these simulations, we only consider the broadcast trajectories used within the ADMM scheme.
Suppose this information is also used by other actors outside the ADMM scheme, e.g., shore control or manned ships.
In that case, we could increase the number of iterations, e.g., $iter_{max}=6$, to avoid misunderstanding of intention from actors outside the ADMM scheme.

\subsection{Intersection crossing scenarios}
In this scenario, two or more ships are sailing towards an intersection.
The situation is set up in such a way that without a collision avoidance action, all ships shall cross the intersection at the same time.
Due to the limited waterway width, a ship cannot change course to avoid collision as would be the case at open sea.
Instead, the expected action (for the give-way ship) is to reduce speed.
Accordingly, the pair of control parameters $K_y$ and $K_u$ is chosen as $K_y =5$ and $K_s = 2\times 10^{-2}$.

In intersection crossing scenarios, it is not easy to visualize the safety area of ships, especially when there is more than two ships.
Therefore we introduce the safety index $\epsilon_i$ to evaluate the safety performance of ship $i$, and is defined as follows:
\begin{align*}
		\epsilon_{i}&= \min_{j\in \Mc \backslash \{ i \} }\left\{\max\{|\ix{x}{i} - \ix{x}{j}|-d^x_i,|\ix{y}{i} - \ix{y}{j}|-d^y_i\}\right\}.
\end{align*}
Here, $d^x_i=51.5$ and $d^y_i=8.6$ are, respectively, half the length and width of the safety area that are illustrated by the dash rectangle in Fig. \ref{ho}.
When $\epsilon_i >0$, it means that there is no other ship in the safety area of ship $i$ and there is no risk of collision at that time.
When $\epsilon_i\leq 0$, it is likely that a collision will happen.
\subsubsection{Intersection crossing of two ships}\label{sec:sim-headon}
\begin{figure*}[!t]
	\centering
	\subfloat[\label{IC21-1}]{
		\centering
		\includegraphics[width=0.3\linewidth]{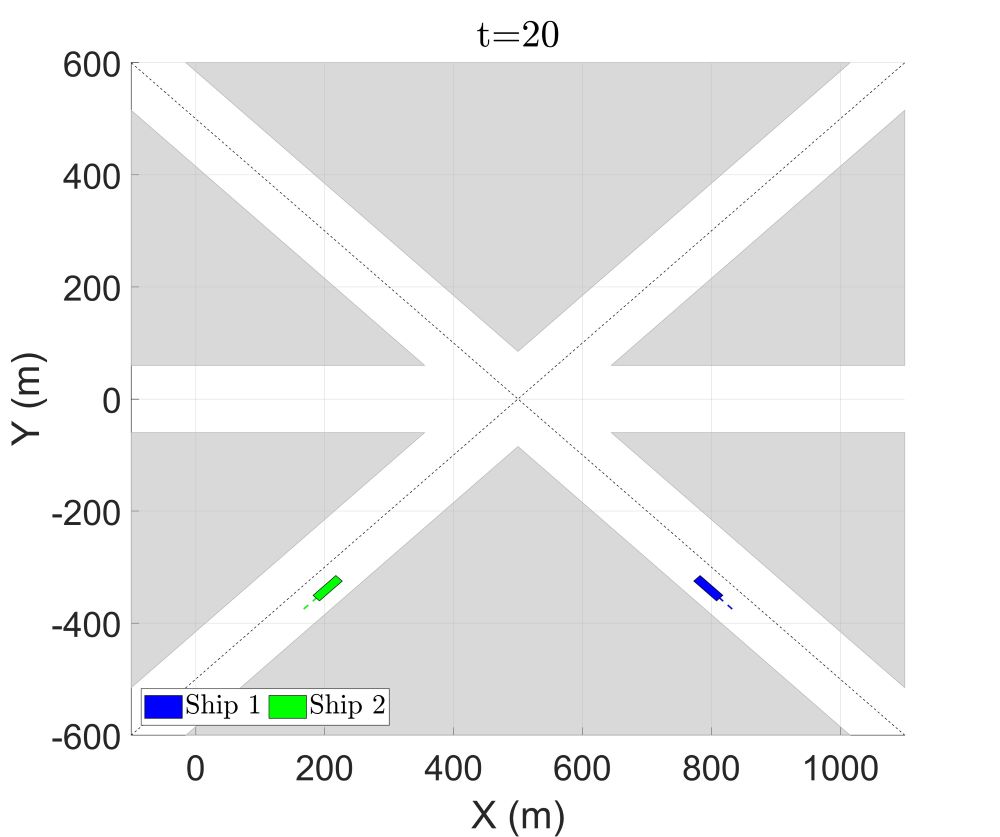}}
	\centering
	\subfloat[\label{IC21-2}]{
		\centering
		\includegraphics[width=0.3\linewidth]{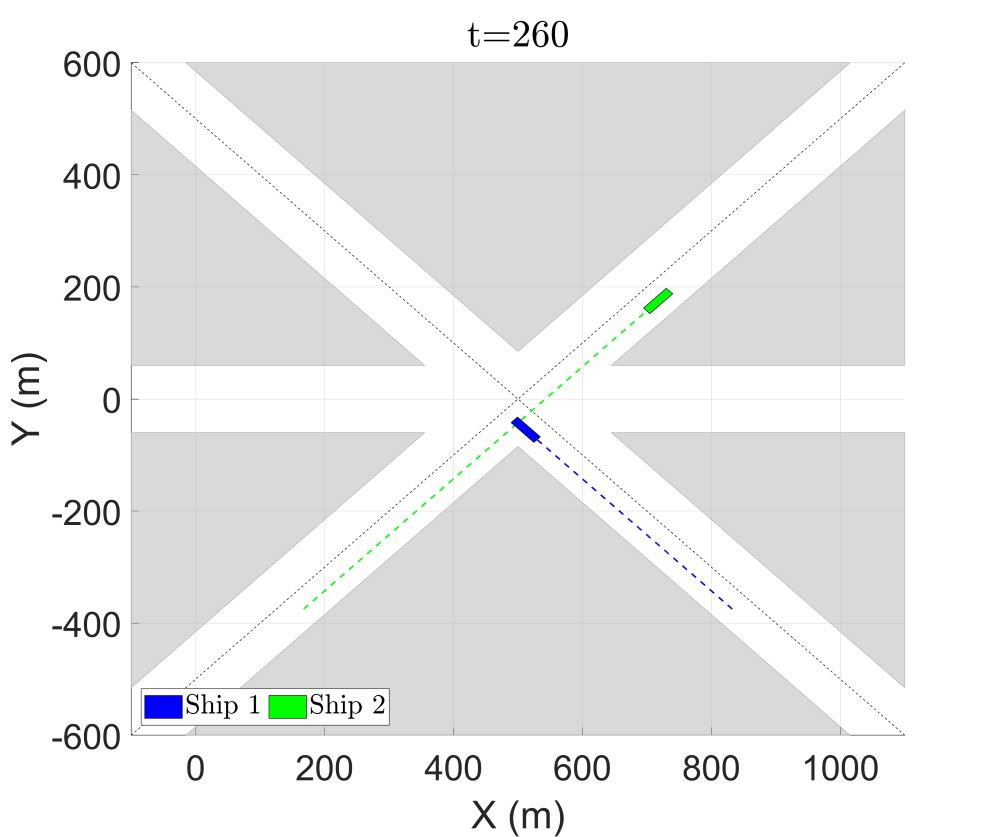}}	
	\centering
	\subfloat[Safety indices of ships\label{IC21-4}]{
		\centering
		\includegraphics[width=0.3\linewidth]{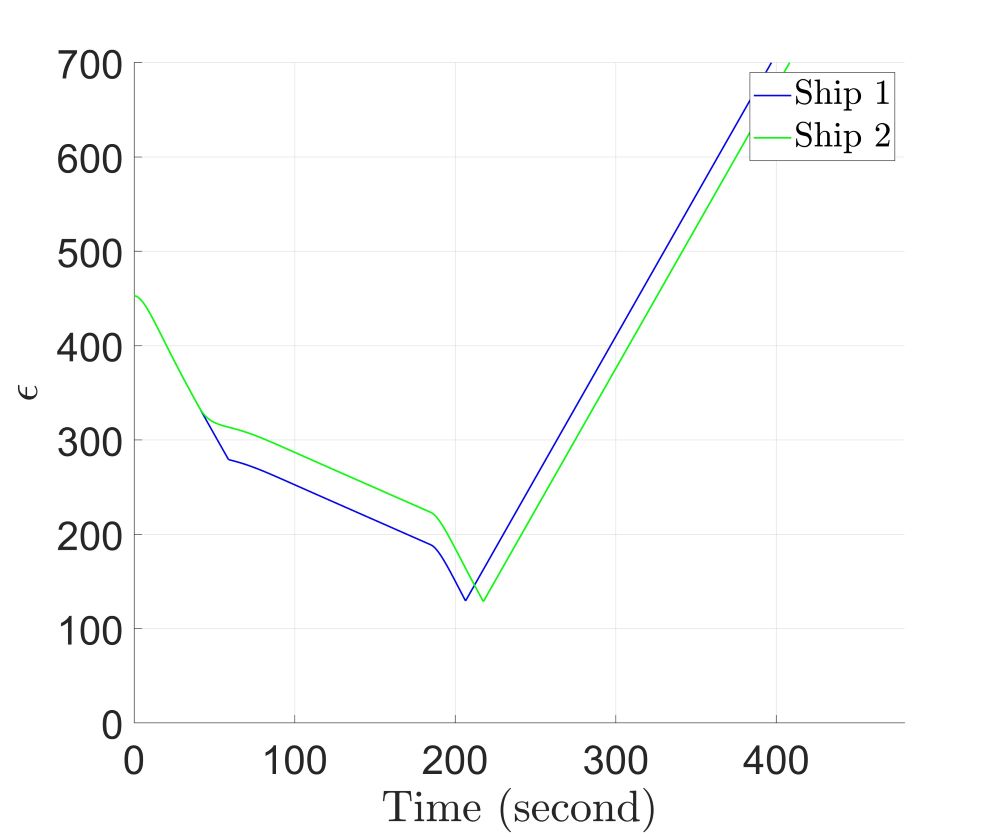}}
	\caption{Intersection crossing between 2 ships. Ship 2 has stand-on priority.}
	\label{IC21}
\end{figure*}
\begin{figure*}[!t]
	\centering
	\subfloat[\label{IC22-1}]{
		\centering
		\includegraphics[width=0.3\linewidth]{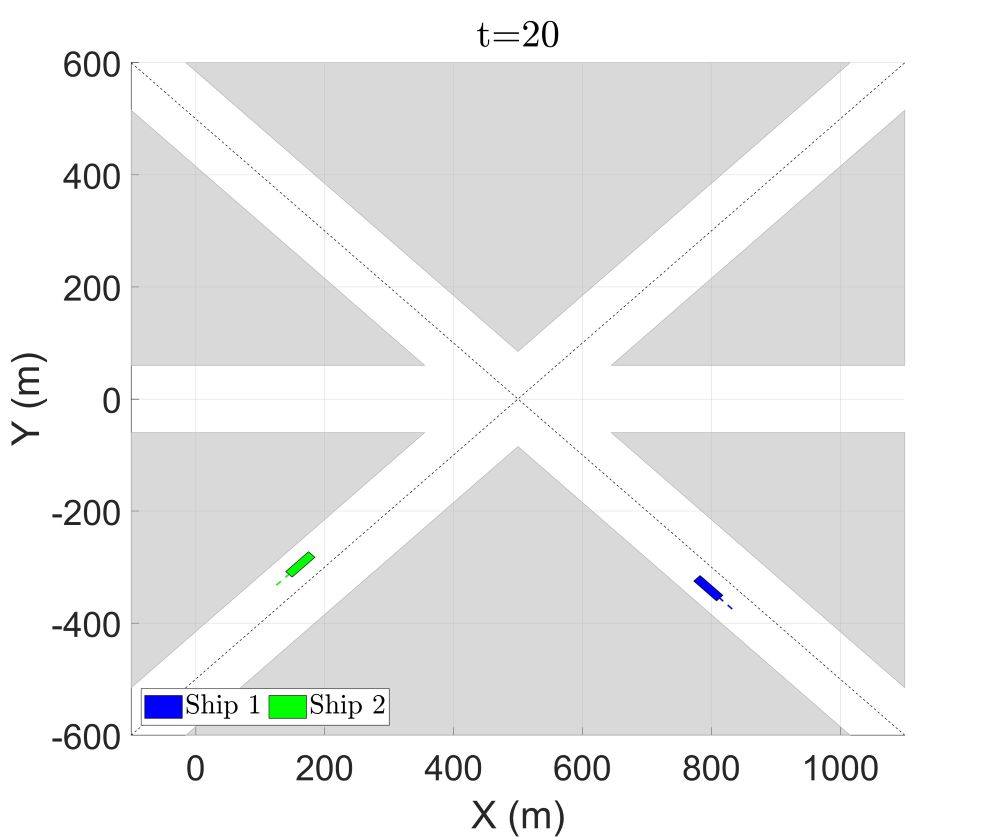}}
	\centering
	\subfloat[\label{IC22-2}]{
		\centering
		\includegraphics[width=0.3\linewidth]{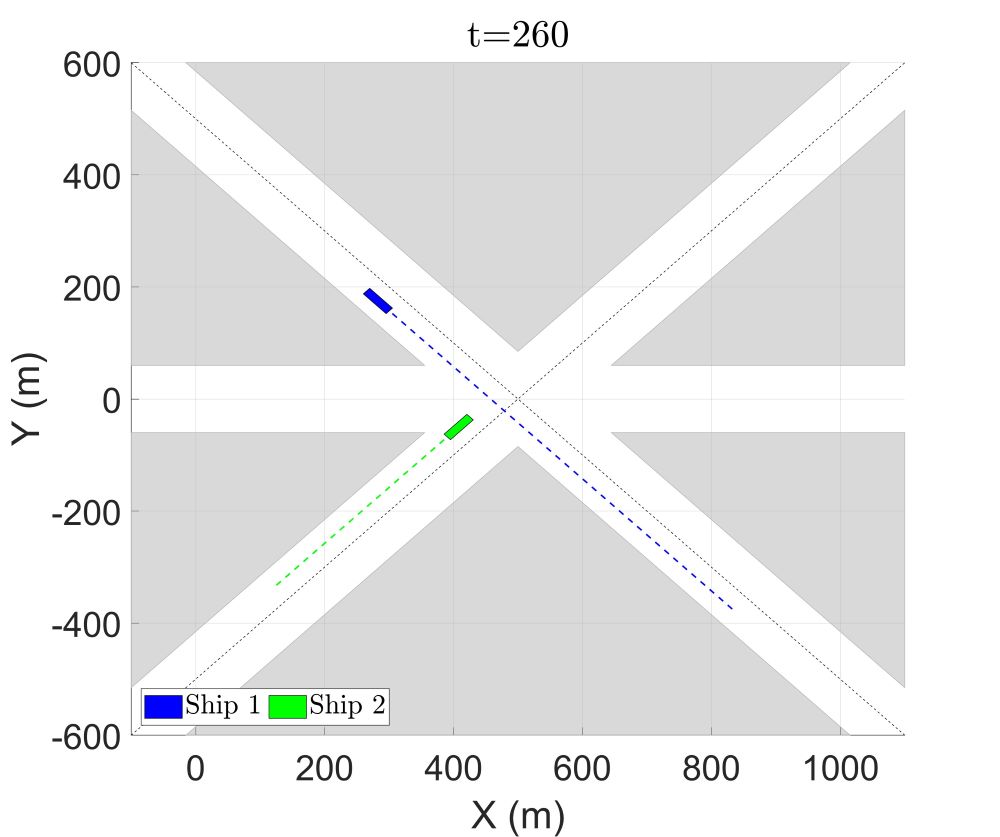}}
	\centering
	\subfloat[Safety indices of ships\label{IC22-4}]{
		\centering
		\includegraphics[width=0.3\linewidth]{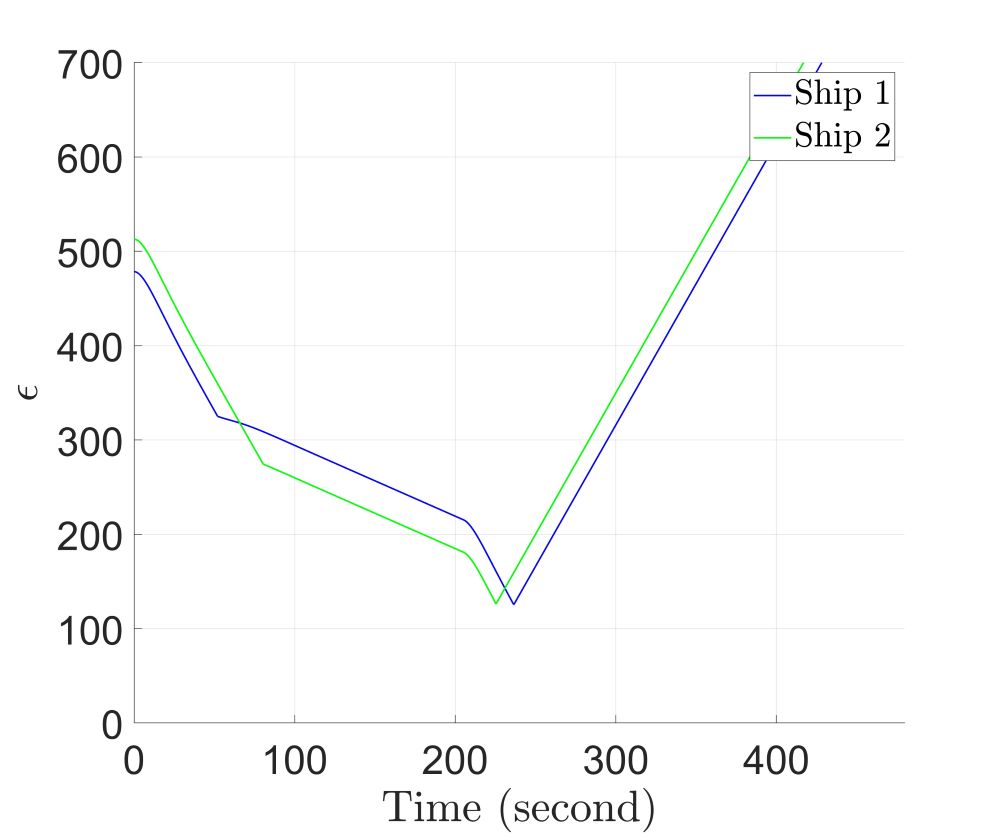}}
	\caption{Intersection crossing between 2 ships. Ship 1 has stand-on priority.}
	\label{IC22}
\end{figure*}
%
%
This is the situation in which two ships cross each other. 
One ship has to give-way to the other by reducing speed.
Following the situation shown in Fig. \ref{IC21-1}, ship 2 is sailing on the starboard side of the waterway while ship 1 is sailing on the port side.
Therefore, in this case, ship 1 is the give-way ship and makes the first collision avoidance decision.
As a result, ship 1 reduces speed and waits for ship 2 to pass the intersection (see Fig. \ref{IC21-2}).
Fig. \ref{IC22} shows a similar scenario, but this time, none of the two ships sails on the starboard side of the waterway.
Therefore, ship 2 has to give way to ship 1 as ship 1 comes from the starboard side of ship 2 (see Fig. \ref{IC22-2}).
\subsubsection{Intersection crossing of more than two ships}\label{sec:sim-ic}
\begin{figure*}[!t]
	\centering
	\subfloat[\label{IC3-1}]{
		\centering
		\includegraphics[width=0.3\linewidth]{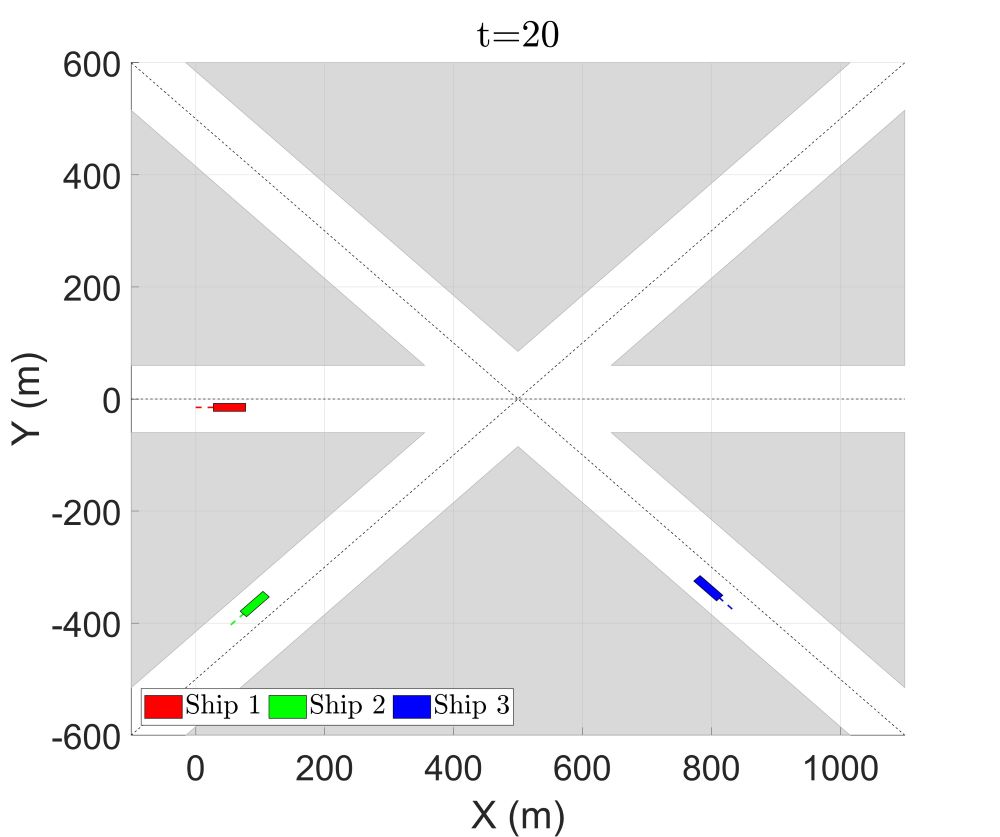}}
	\centering
	\subfloat[\label{IC3-3}]{
		\centering
		\includegraphics[width=0.3\linewidth]{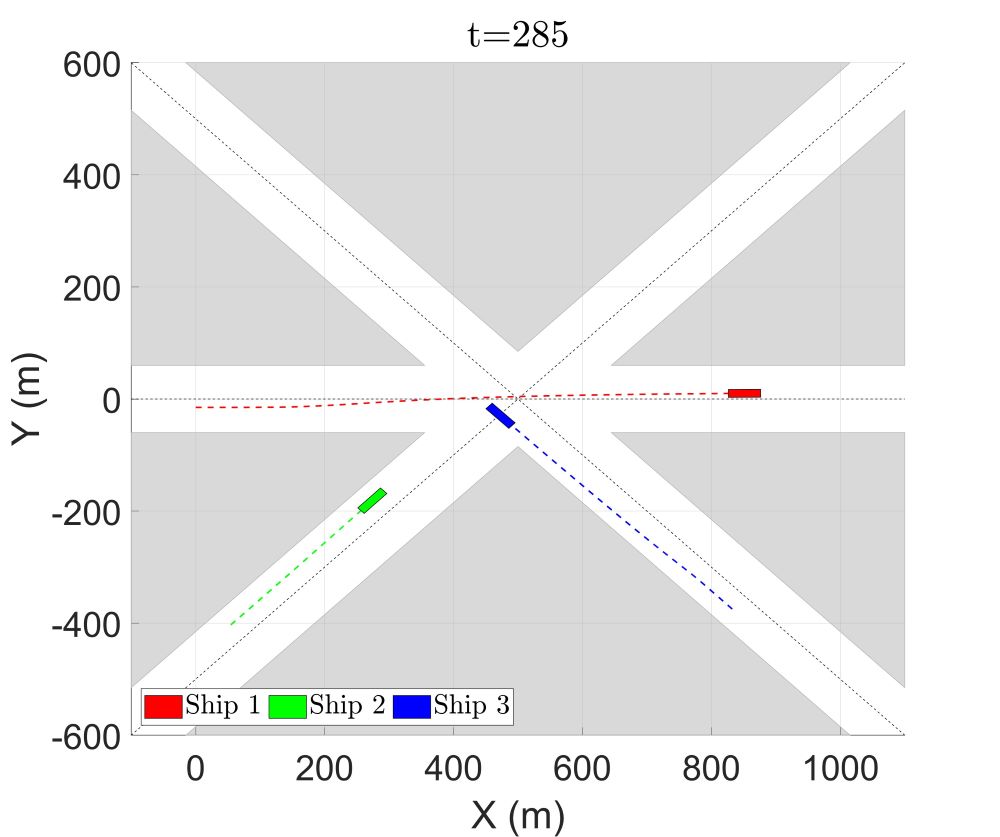}}
    \hfill
	\centering
	\subfloat[\label{IC3-4}]{
		\centering
		\includegraphics[width=0.3\linewidth]{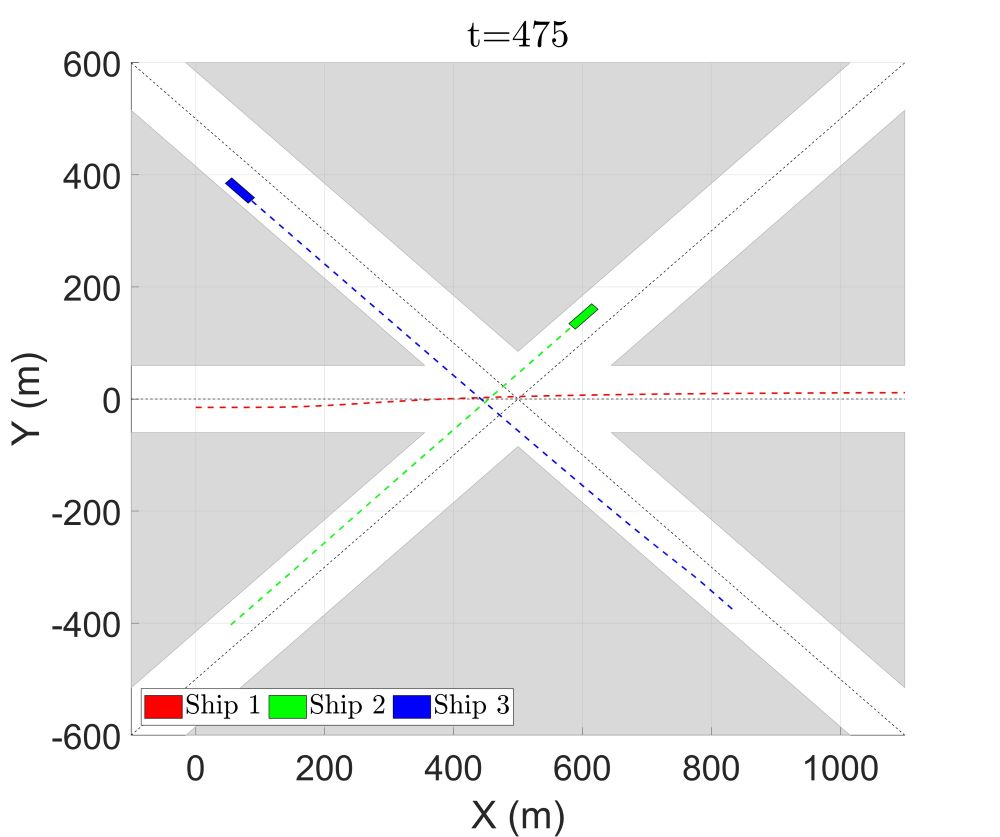}}
    \subfloat[Safety indices of ships\label{IC3-5}]{
		\centering
		\includegraphics[width=0.3\linewidth]{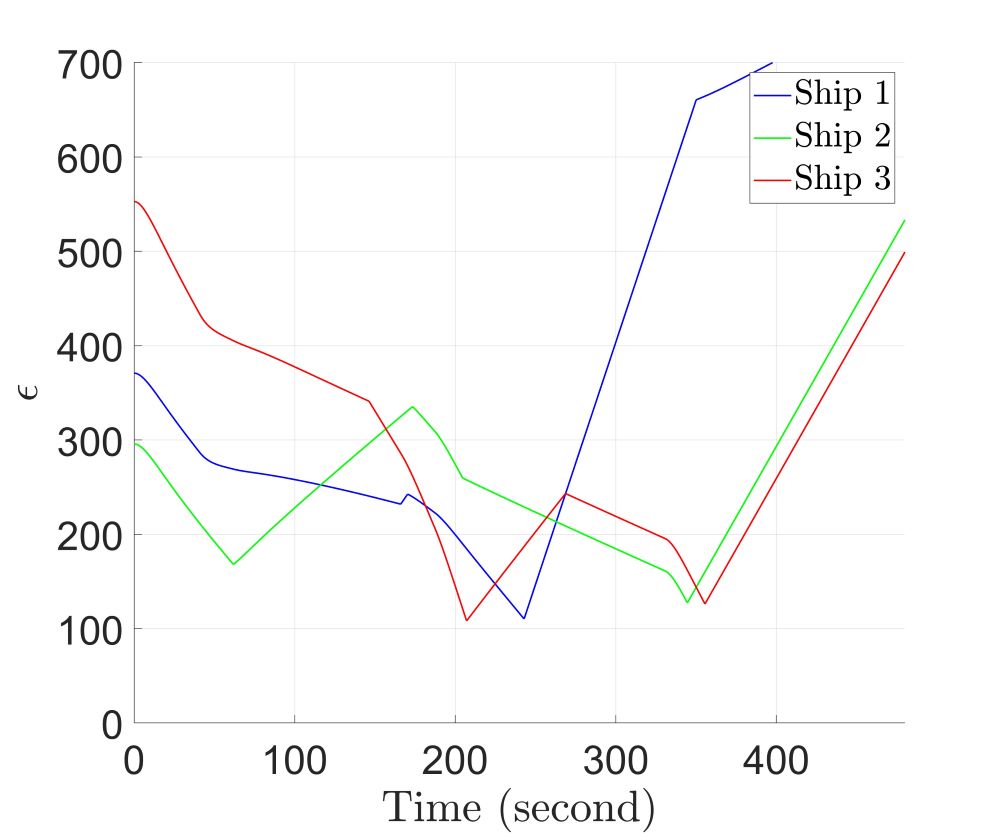}}
	\caption{Intersection crossing between 3 ships.}
	\label{IC3}
\end{figure*}
\begin{figure*}[!t]
	\centering
	\subfloat[\label{IC42-1}]{
		\centering
		\includegraphics[width=0.3\linewidth]{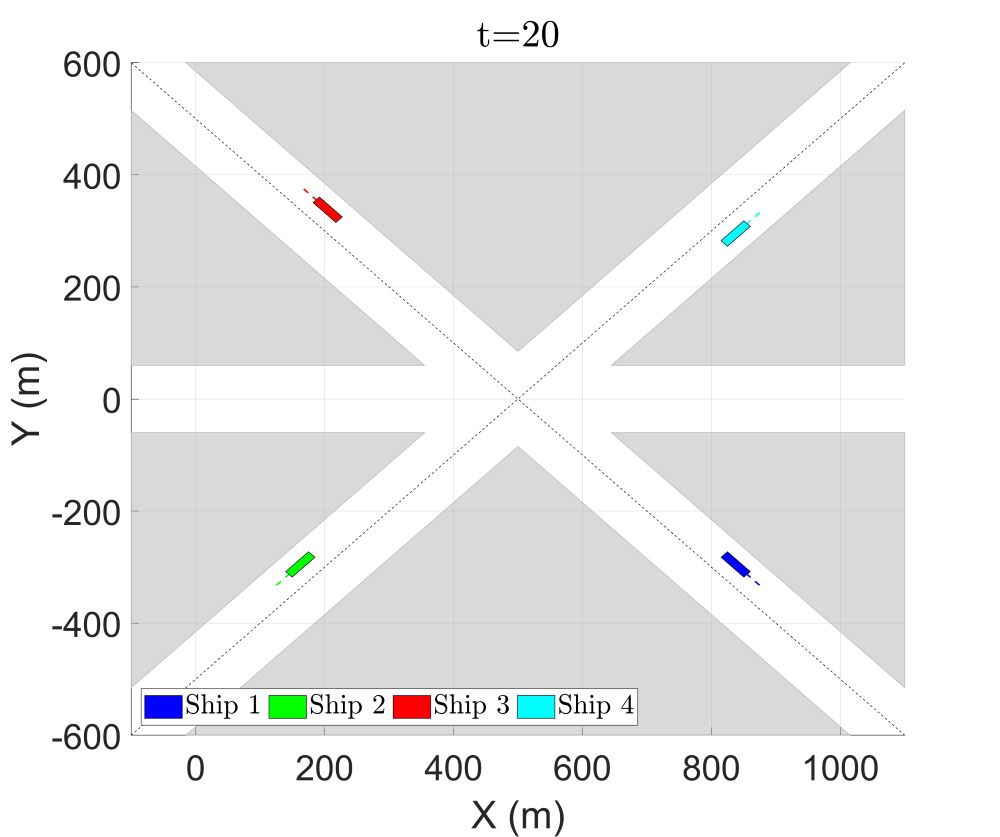}}
	\centering
	\subfloat[\label{IC42-3}]{
		\centering
		\includegraphics[width=0.3\linewidth]{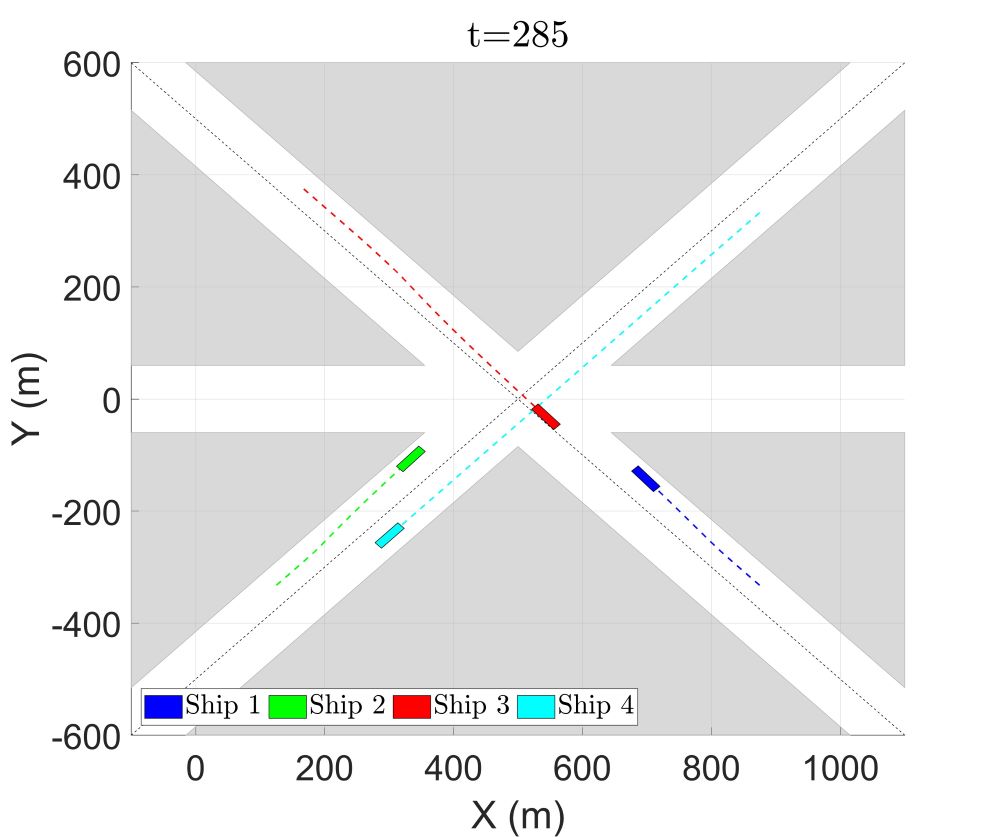}}
	\centering
    \hfill
	\subfloat[\label{IC42-4}]{
		\centering
		\includegraphics[width=0.3\linewidth]{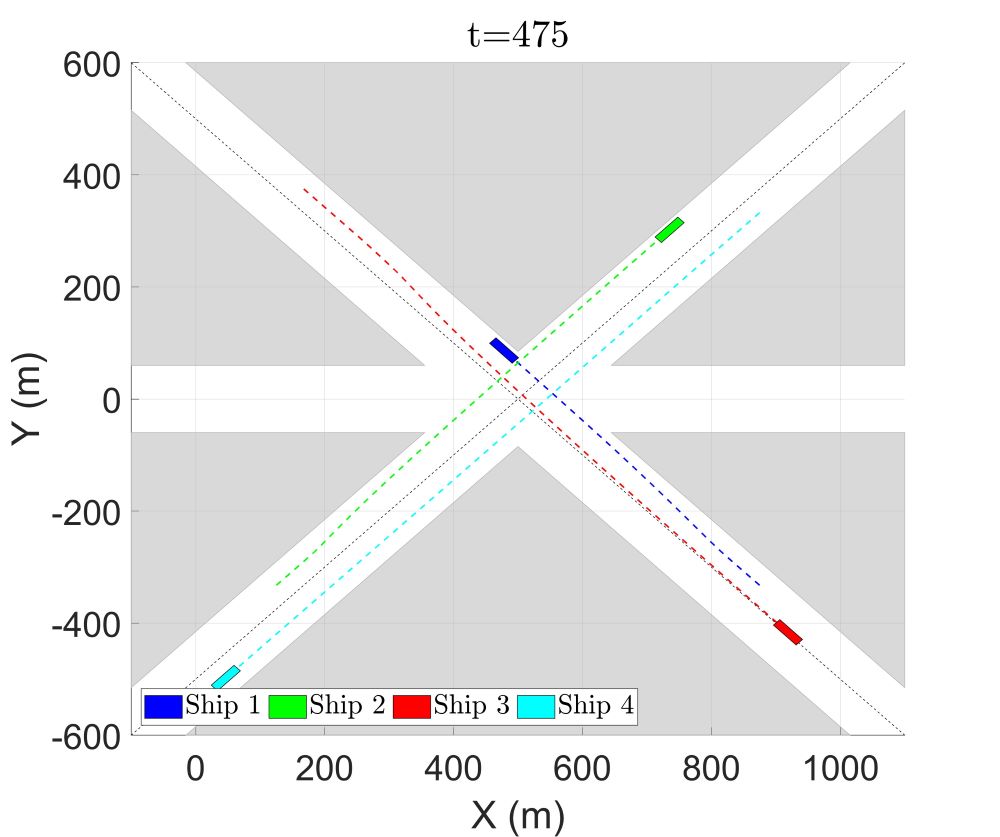}}
        \centering
	\subfloat[Safety indices of ships\label{IC42-5}]{
		\centering
		\includegraphics[width=0.3\linewidth]{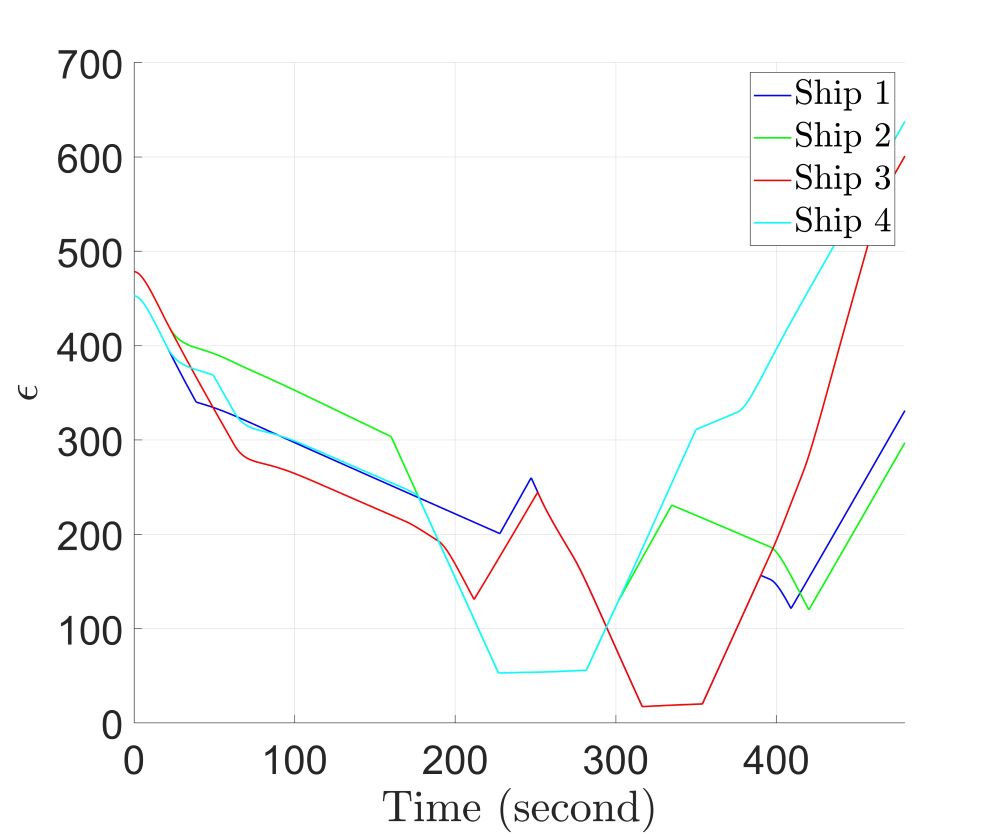}}
	\caption{Intersection crossing between 4 ships.}
	\label{IC42}
\end{figure*}
\begin{figure*}[!t]
	\centering
	\subfloat[Deadlock situation.]{
		\centering
		\includegraphics[width=0.3\linewidth]{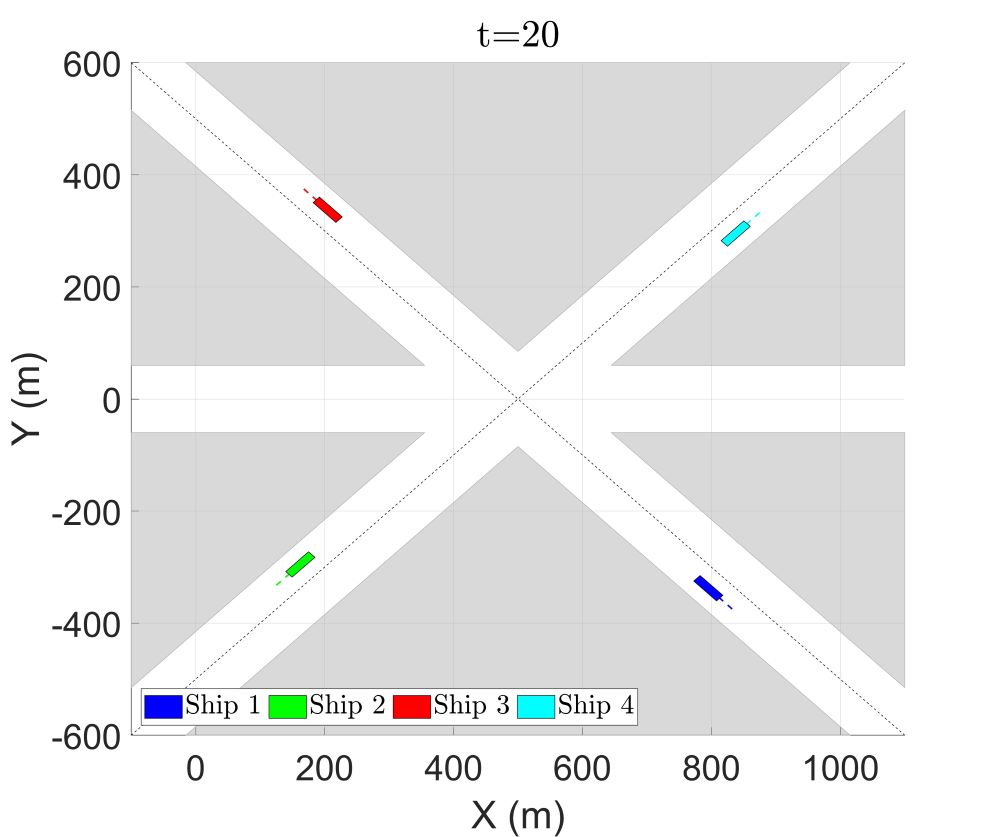}
		\label{IC41-1}}
	\centering
	\subfloat[Result without deadlock detection.]{
		\centering
		\includegraphics[width=0.3\linewidth]{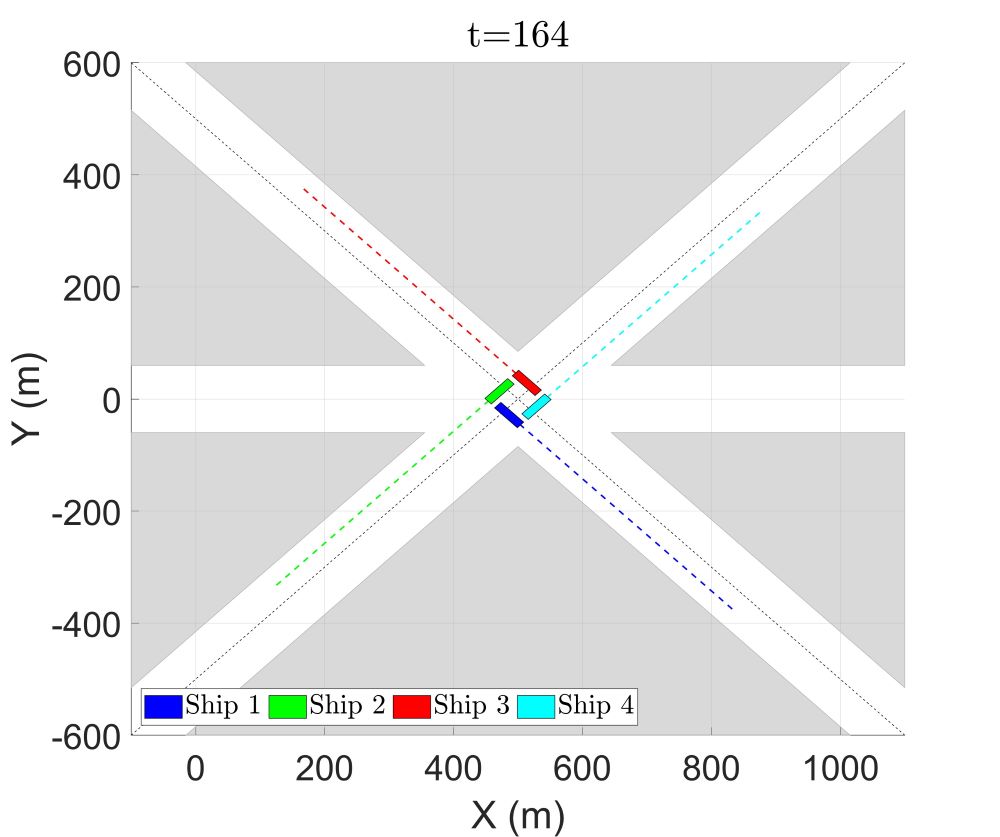}
        \label{IC41-2}}
	\centering
	\subfloat[Result with deadlock detection.]{
		\centering
		\includegraphics[width=0.3\linewidth]{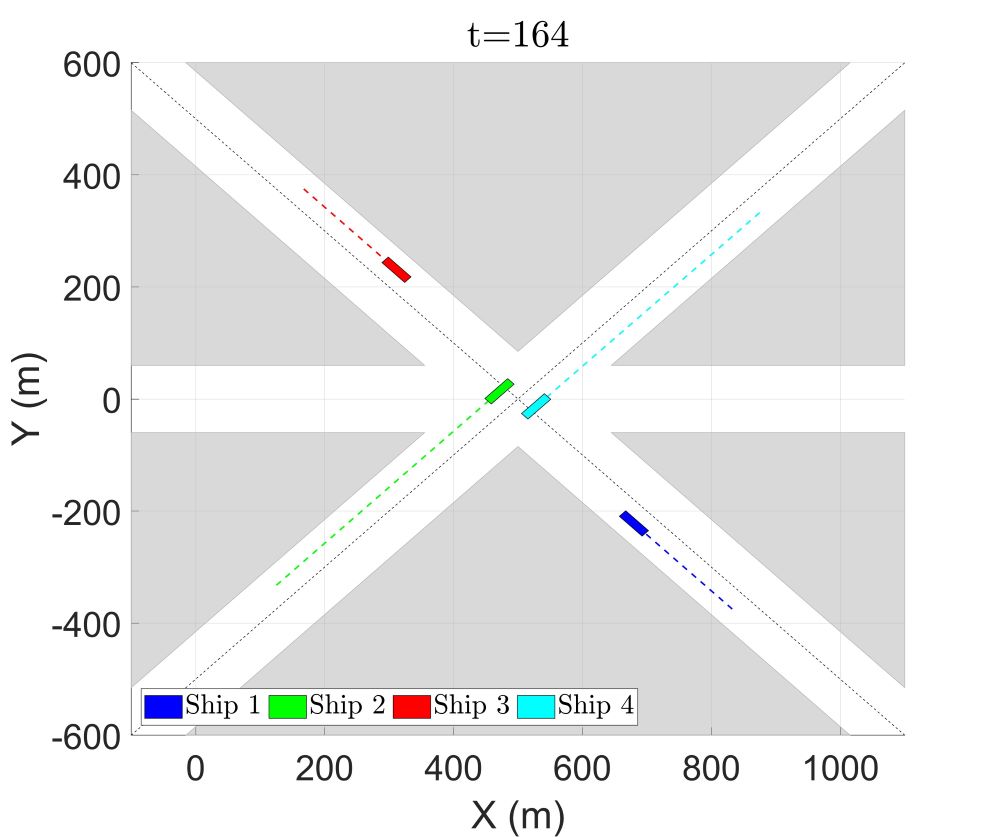}
        \label{IC41-3}}
	\caption{Intersection crossing between 4 ships in a deadlock situation.}
	\label{IC41}
\end{figure*}
%
%

Fig. \ref{IC3} shows a situation with three ships crossing the intersection.
According to the traffic rules, ship 1 can stand on, ship 3 gives way to ship 1, and ship 2 gives way to ships 1 and 3.
The results show that the solution given by the proposed algorithm follows the traffic regulation.
Firstly, as shown in Fig. \ref{IC3-3}, ships 2 and 3 reduce speed to give way to ship 1.
When ship 1 safely crosses the intersection, ship 2 keeps a low speed to give way to ship 3.
In our experiments, we can increase $K_y$ to reduce the unnecessary course change (as ship 1 in Fig. \ref{IC3-3}).
However, increasing $K_y$ also reduces the ability to avoid collision with other ships coming from the opposite direction (head-on situation).
Therefore, $K_y$, in this case, is chosen in such a way that the ship will prioritize reducing speed but can change course when reducing speed cannot resolve the collision avoidance problem.
%

When more than two ships cross the intersection simultaneously, the collision avoidance problem becomes more complex.
In complex situations, ships may disregard the give-way (or stand-on) roles and pursue decisions based on mutual benefits.
This is because, in our C-CAS framework, the traffic rules are not hard constraints that force ships to follow.
The situation in Fig. \ref{IC42} is an example of a complex situation in which ships are involved in the head-on and crossing situation simultaneously.
As shown in Fig. \ref{IC42-1}, four ships are heading toward the intersection, in which ship 1 and 3 are also in a head-on situation.
Ship 1 has the stand-on priority and is supposed to be given way by three other ships.
However, the collision risk between ship 1 and ship 3 caused these two ship to initially reduce their speed.
This gives the opportunity for ship 4 to safely cross the intersection first.
Because $K_y$ is large, then follows the experiment in Section \ref{HO-scenarios}, both ship 1 and 3 change course to avoid collision and cross the intersection.

In the last experiment, we evaluate the ability to resolve the deadlock situation as mentioned in Section \ref{sec:comm-deadlock}.
Results are given in Fig. \ref{IC41}.
We assume that ship 1 is the smallest ship and make the first decision.
Without the detection of a deadlock situation, the an accident could happen as shown in Fig. \ref{IC41-2}.
However, with the proposed method in Section \ref{sec:comm-deadlock}, four ships safely cross the intersection (see Fig. \ref{IC41-3}).
Furthermore, although being mentioned in Section \ref{sec:comm-prot} that a deadlock situation could arise when Algorithm \ref{alg:protocol} fails to fulfill the condition of consistency, this situation have never been encountered in our simulations experiments.

\subsection{Choice of $K_y$ and $K_s$ based on traffic scenarios}
In Section \ref{sec:sim-headon} and Section \ref{sec:sim-ic}, we used the same control parameters for both scenarios, except for two parameters, $K_y$ and $K_s$.
The reason for the difference between $K_y$ and $K_s$ in the two scenarios is that we expect different ship behavior in each scenario.
We can use the same parameters ($K_y$ and $K_s$) for two scenarios, but it will result in the unfavorable behaviors of ships.
For example, if we choose $K_y=2 \times 10^{-2}$ in intersection crossing scenarios, ships tend to make more unnecessary course changes before reducing speed.
In practice, a ship can automatically switch between two sets of parameters based on the situation determination in steps \ref{alg2:cross-situation} and \ref{alg2:ho-situation} of Algorithm \ref{alg:protocol}.

\section{Conclusions and future research} \label{sec:cons}
This paper presented a two-layer framework for distributed collaborative collision avoidance of autonomous ships in inland waterways.
We used the two-layer C-CAS framework to help ships to better comply with traffic regulations.
By implementing a decision-making order based on priority, we separated the task of traffic rule compliance from collision avoidance control.
Our proposed framework allows ships to follow different traffic rules without modifying the control algorithm.
Furthermore, we introduced a ADMM-based DMPC algorithm that is designed for inland waterway traffic.
The simulation experiments illustrate the performance of our proposed algorithm in various typical traffic scenarios.

In future research, we will focus on increasing the resilience of our algorithm.
For example, this paper did not consider the case when one ship fail to make decision or cannot broadcast its decision.
This situation could harm the stability of the network and need further investigation.
An asynchronous communication scheme will also be developed to make the algorithm better suited for practical applications.
\appendix
\section{Proof of Theorem \ref{theorem:converge}} \label{proof}
Let us introduce the following notation and definitions that are used in this proof.
We denote $\textbf{range}(M)$ as the range (column space) of matrix M. 
The domain of an extended-real-valued function $f: \Rbb^n \rightarrow \overbar{\Rbb}$ is $\textbf{dom} f:= \left\{x \in \Rbb^n~|~f(x) < \infty \right\}.$
We also uses the notion of lower semicontinuous function (lsc), image function, and Lipchitz continuous gradient defined as follows:
\begin{definition}[lower semicontinuous function]
	A function $f: X \rightarrow \overbar{\Rbb}$ is called lower semicontinuous (lsc) at point $x_0 \in X$ if $\lim \inf_{x \rightarrow x_0} f(x) = f(x_0)$. Furthermore, $f$ is called lsc if $f(x)$ is lsc at every point $x_0 \in \textbf{dom}f$
\end{definition}
\begin{definition}[image function]
	Given $f: \Rbb^n \rightarrow \overbar{\Rbb}$ and $M \in \Rbb^{m \times n}$. Then the image function $(Mf): \Rbb^m \rightarrow [-\infty,+\infty]$ is defined as $(Mf)(\epsilon) := \inf_{x \in \Rbb^n}\left\{f(x)~|~Mx=\epsilon \right\}$.
\end{definition}
\begin{definition}[Lipchitz continuous gradient]
	A differentiable function $h$ is said to have Lipschitz continuous gradient with constant $L_h>0$ (or $L_{h}-\text{smooth}$) on $\textbf{dom}(h)$ if 
	\begin{align}
		||\nabla h(x_1)- \nabla h(x_2)|| \leq L_h ||x_1 -x_2||, ~~ \forall x_1,x_2 \in \textbf{dom}(h).
	\end{align}
\end{definition}

Next, we write the optimization problem of $M$ ships in $\Mc$, where each individual problem follows \eqref{prob2}, in the form of problem \eqref{ref-prob} with:
\begin{align}\label{prob2:newvar}
	\begin{split}
		w &= [\tilde{u}_1^\top, \tilde{p}_{1,1}^\top,...,\tilde{u}_M^\top, \tilde{p}_{M,M}^\top]^\top, \\
		v &= [\bar{p}_{1,1}^\top,...,\bar{p}_{M,M}^\top]^\top,\\
		f(w) &= \sum_{i \in \Mc}\Jc_i(\bpi{i} ,\tilde{u}_i) + \Gc_i(\bpi{i} ,\tilde{u}_i), \\
		g(v) &=0, \\
		A &= I_{M} \otimes \bbm 0_{3(N+1) \times 2N} & I_{3(N+1)}\ebm,\\
        B &= -I_{3M(N+1)},~b=0,
	\end{split}
\end{align}
where $\Gc_i(\tilde{u}_i,\bpi{i})$ is an indicator function that is defined as $\Gc_i(\tilde{u}_i,\bpi{i})=0$ if $[\tilde{u}_i^\top,\bpi{i}^\top]^\top \in \Gbf_i$ and $+\infty$ otherwise.
Problem \eqref{prob2} is a special case of problem \eqref{ref-prob}, in which $g(v)=0$ and $B= -I$.
In this case, the NADMM update \eqref{NADMM} becomes:
\begin{align}\label{NADMM-2}
	\begin{cases}
		z^{+1/2} &= z - \beta(1-\lambda)(Aw - v -b),\\
		w^+      &\in \arg\min \Lc_{\beta}(\cdot,v,z^{+1/2}),  \\
		z^{+}    &= z^{+1/2} +  \beta(Aw - v -b),\\
		v^+      &= Aw^+ -b + \frac{1}{\beta}z^{+}.
	\end{cases}
\end{align}
%

According to \cite[Thm.~5.6]{themelis_20}, Theorem \ref{theorem:converge} holds if Problem \eqref{prob2} satisfies \cite[Asm.~II]{themelis_20}: 
\begin{enumerate}
	\item $f$ and $g$ are proper and lsc.
	\item $A$ is surjective.
	\item $(Af)$ is $L_{(Af)}-\text{smooth}$.
	\item $(Bg)$ is lsc.
\end{enumerate}

Since $g(v) = 0$ then it is trivial that $g$ is proper, lsc, an $(Bg)$ is lsc.
We also have $A$ is surjective because $A$ is full row rank.

Now we prove that $f(w)$ is lsc.
From Problem \eqref{prob2}, and \eqref{prob2:newvar} we also have $f(w) = \sum_{i \in \Mc}\Jc_i(\bpi{i} ,\tilde{u}_i) + \Gc_i(\bpi{i} ,\tilde{u}_i)$. 
Taking into account that $\Gc_i(\bpi{i} ,\tilde{u}_i)$ is the indicator function of a closed set then $\Gc_i(\bpi{i} ,\tilde{u}_i)$ is $lsc$ \cite[Prop.~1.2.2]{hiriart-urruty_convex_1993}.
Besides, $\Jc_i(\bpi{i} ,\tilde{u}_i)$ is sum of continuous functions hence it is $lsc$.
Therefore, $f(w)$ is $lsc$.

Following to \cite[Thm.~5.13]{themelis_20}, $(Af)$ is $L_{(Af)}-\text{smooth}$ if there exist $L_{(Af)}>0$ such that
\begin{align}\label{L-con}
	-L_{(Af)} ||A(w_1-w_2)||^2 &\leq \langle \nabla f(w_1) - \nabla f(w_2),w_1-w_2\rangle \non\\
    &\leq L_{(Af)}||A(w_1-w_2)||^2
\end{align}
whenever $\nabla f(w_1), \nabla f(w_2) \in \textbf{range} (A^\top)$. 

From \eqref{prob2:newvar}, we have $A = I_{M} \otimes \bbm 0_{3(N+1) \times 2N} & I_{3(N+1)}\ebm$, in which case $\nabla f(w)\in \textbf{range} (A^\top)$ if and only if $\tilde{u}_i =0, \forall i \in \Mc$ and $\Gc_i = 0, \forall i \in \Mc$. 
Then $f(w) = \sum_{i \in \Mc} \Jc_i^{ca}(\bpi{i})$ is Lipschitz continuous gradient function because $R_{ij}(\cdot)$ is $L_{R}-\text{smooth}$.
This implies that exist $L_f$ such that
\begin{align*}
	||\nabla f(w_1)- \nabla f(x_2)|| \leq L_f ||w_1 -w_2||.
\end{align*}
Using the Cauchy-Schwartz inequality we obtain
%
\begin{align*}
	L_f ||w_1 -w_2||^2 &= L_f ||A(w_1 -w_2)||^2 \\
    &\geq|\langle \nabla f(w_1) - \nabla f(w_2),w_1-w_2\rangle| \\
    &\geq |\nabla f(w_1)- \nabla f(x_2)|. |w_1 -w_2|,
\end{align*}
for all $\nabla f(w)\in \textbf{range} (A^\top)$.
Then there exist  $L_{(Af)}$ satisfy \eqref{L-con}. \qedsymbol

\bibliographystyle{IEEEtran}
\bibliography{mybibfile,MyEndnoteLibrary}

\end{document}